\documentclass{article}

\usepackage{arxiv}

\usepackage[utf8]{inputenc} 
\usepackage[T1]{fontenc}    
\usepackage{url}            
\usepackage{booktabs}       
\usepackage{amsfonts}       
\usepackage{nicefrac}       
\usepackage{microtype}      
\usepackage{multirow}
\usepackage{amsmath}
\usepackage{graphicx}
\usepackage{subfigure}
\usepackage{lipsum}
\usepackage{color}
\usepackage{footnote}
\usepackage[LAE,T1]{fontenc}
\usepackage[arabic,english]{babel}
\usepackage{authblk}

\usepackage[misc]{ifsym}
\usepackage{fancyvrb,cprotect}

\title{SenWave: Monitoring the Global Sentiments under the COVID-19 Pandemic}

\author[a]{Qiang Yang}
\author[a]{Hind Alamro}
\author[a]{Somayah	Albaradei}
\author[a]{Adil Salhi}
\author[b]{Xiaoting Lv}
\author[a]{Changsheng Ma}
\author[a]{Manal Alshehri}
\author[c]{Inji Jaber}
\author[a]{Faroug Tifratene}
\author[a,b]{Wei Wang}
\author[a]{Takashi Gojobori}
\author[d]{Carlos M. Duarte}
\author[a]{Xin Gao}
\author[a]{Xiangliang Zhang{\tiny \Letter}}  
\affil[a]{Computer, Electrical and Mathematical Sciences and Engineering (CEMSE) Division and Computational Biosciences Research Center (CBRC), King Abdullah University of Science and Technology (KAUST) \authorcr Email: \{qiang.yang, hind.alamro, somayah.albaradei, adil.salhi, changsheng.ma, manal.alshehri, faroug.tifratene, wei.wang, takashi.gojobori, xin.gao, xiangliang.zhang\}@kaust.edu.sa}
\affil[b]{School of Computer Science and Information Technology, Beijing Jiaotong University, China \authorcr Email: xiaotinglvtt@gmail.com}
\affil[c]{IT department, King Abdullah University of Science and Technology (KAUST) \authorcr Email: inji.jaber@kaust.edu.sa}
\affil[d]{Red Sea Research Center (RSRC) and Computational Biosciences Research Center (CBRC), King Abdullah University of Science and Technology (KAUST) \authorcr Email: carlos.duarte@kaust.edu.sa}

\begin{document}
\maketitle

\begin{abstract}
Since the first alert launched by the World Health Organization (5 January, 2020), COVID-19 has been spreading out to over 180 countries and territories. As of June 18, 2020, in total, there are now over 8,400,000 cases and over 450,000 related deaths. 
This causes massive losses in the economy and jobs globally and confining about 58\% of the global population. 
In this paper, we introduce SenWave, a novel sentimental analysis work using  105+ million collected tweets and Weibo messages to evaluate the global rise and falls of sentiments during the COVID-19 pandemic. To make a fine-grained analysis on the feeling when we face this global health crisis, we annotate 10K tweets in English and 10K tweets in Arabic in 10 categories, including \emph{optimistic, thankful, empathetic, pessimistic, anxious, sad, annoyed, denial, official report}, and \emph{joking}. 
We then utilize an integrated transformer framework, called \emph{simpletransformer}, to conduct multi-label sentimental classification by fine-tuning the pre-trained language model on the labeled data.
Meanwhile, in order for a more complete analysis, we also translate the annotated English tweets into different languages (Spanish, Italian, and French) to generated training data for building sentiment analysis models for these languages.
SenWave thus reveals   the sentiment of global conversation in six different languages on COVID-19 (covering English, Spanish, French, Italian, Arabic and Chinese), followed the spread of the epidemic. The conversation showed a remarkably similar pattern of rapid rise and slow decline over time across all nations, as well as on special topics like the herd immunity strategies, to which  the global conversation reacts strongly negatively. Overall, SenWave shows that optimistic and positive sentiments increased over time,  foretelling a desire to seek, together, a reset for an improved COVID-19 world.
\end{abstract}


\section{Introduction}
Since the outbreak of coronavirus, it has affected more than 180 countries where massive losses in the economy and jobs globally and confining about 58\% of the global population are caused. Many people have been forced to work or study from home under pandemic. The research on people's feelings is   essential for keeping mental health and informed about Covid-19. Social medias (e.g., Twitter, Weibo) have played a major role in expressing people's feelings and attitudes towards Covid-19. We thus target on building a system named SenWave to monitor the global sentiments under the COVID-19 Pandemic by deep learning powered sentiment analysis.

Sentiment analysis has been widely researched in the field of natural language processing~\cite{anees2020survey,zhang2018deep,kharde2016sentiment,medhat2014sentiment}. 
%
%
%
~
Most of the current sentimental analysis tasks   usually consider the coarse-grained emotion labels like \emph{positive, neutral}, and \emph{negative} for reviews/comments of books/products/movies, or five values to indicate the degree of emotions with the ranking score from $1$ to $5$. However, the feelings of people in pandemic are much more complicated than the sentiments in movie reviews and product comments ect. For instance,   people may  negatively feel \emph{angry} and \emph{sad} since Covid-19 leads to the increasing number of deaths and unemployment, while others may feel \emph{optimistic} because of the medical supplies and medical assistance for the people in need.
Therefore, we need to define the fine-grained labels to better understand the impact of the health crisis on sentiment. 

However, there are no appropriate and sufficient annotated data to support the building of our SenWave by training deep learning sentiment classifiers. One non-Covid-19 tweet sentimental analysis dataset is avaliable in ~\cite{SemEval2018Task1} with the labeled 7724 English tweets, 2863 Arabic tweets, and 4240 Spanish tweets (in total 14, 827), which is a benchmark dataset labeled in 11 categories, \emph{anger, anticipation, disgust, fear, joy, love, optimism, pessimism, sadness, surprise}, and \emph{trust}. Initially, we tried to use it as our training data for building a Covid-19 sentiment classifier. However, the sentimental results were not suitable for Covid-19 analysis. For example, very few tweets are in the categories \emph{joy, love} and \emph{trust}. Besides, many tweets of \emph{official} reports were classified into inappropriate categories as well as the tweets making \emph{jokes} and 
\emph{denying} conspiracy theories. Actually, these kinds of labels are essential in expressing opinions and attitudes according to our observations.

Due to the lack of appropriate training data, the Covid-19 tweet sentiment analysis has been conducted mostly based on engineered features or conventional bag-of-words derived representations not touching the deep learning models, either in unsupervised or supervised ways but with limited training data (e.g., only 5K in  \cite{kleinberg2020measuring}). 
Yasin et al. built a real-time Covid-19 tweets analyzer using LDA model in USA data on \emph{positive, neutral}, and \emph{negative} \cite{kabir2020coronavis}. Jia et al. used LDA model and NRC Lexicon on the English tweets to predict one of emotions from (\emph{anger, anticipation, fear, surprise, sadness, joy, disgust} and \emph{trust})~\cite{xue2020machine}. Mohammed et al. employed naïve Bayes model to predict Saudis’ attitudes towards Covid-19 preventive measures on the (\emph{postive, negative}, or \emph{neutral}). Caleb et al. used logistic regression classifier  with linguistic features, hashtags and tweet embedding to identify anti-Asian hate and counterhate text~\cite{ziems2020racism}. However, these methods are either limited by the sentimental dictionary and its availability or in lack of the deep understanding of tweets semantics. Even if the supervised methods are used like naïve Bayes, they still cannot satisfy the real case (multi-class classification) since emotions in Covid-19 can be a mixture of  multiple emotions 
(multi-label classification).

In order to solve above problems, on one hand, we collected over 105+ million tweets covering six different languages, including English, Spanish, French, Arabic and Italian from  March 1, 2020. 
On the other hand, to meet the requirement of fine-grained sentiment analysis target for Covid-19, we annotated 10,000 tweets in English and 10,000 tweets in Arabic in 10 categories, including \emph{optimistic, thankful, empathetic, pessimistic, anxious, sad, annoyed, denial, official}, and \emph{joking}. Each tweet was annotated by at least three experienced annotators in the corresponding language under strict quality control. We allowed one tweet to be annotated by more than one category, to support the multi-label analysis. To analyze over 1 million Chinese weibo  of COVID-19 posts, we construct a set of  21,173 Weibo posts labeled in 7 sentimental categories, such as \emph{anger, disgust, fear, optimism, sadness, surprised}, and \emph{gratitude}.
These 41K labeled datasets and the over 106+ million unlabeled tweets and weibo posts compose our dataset  studied by SenWave. More details about dataset are shown in Sec.~\ref{sec:dataset}.
We will make available the data and the SenWave implementation in public for supporting other social impact analysis of Covid-19 and fine-grained sentiment analysis.
To comply with Twitter’s Terms of Service, we will only publicly release the Tweet IDs for unlabeled data and limited number of tweet texts for labeled data (totally 20K) for non-commercial research use. To the best of our knowledge, this is the largest labeled Covid-19 sentimental analysis dataset with the fine-grained labels.
%

We summarize our main contributions in this paper as follows:
\begin{itemize}
\item We constructed the so far the largest fine-grained annotated Covid-19 tweets dataset (10K for English tweets and 10K for Arabic tweets) in 10 sentiment categories, which  help to facilitate the studies of social impact of Covid-19 and other fine-grained analysis tasks in research community. 

\item We share a large set of Covid-19 tweets IDs collected since Mar 1, 2020, in five languages accumulated over 105+ million tweets, which will be continuously updated. 

\item We report the usability of the labeled Covid-19 tweets by first evaluating the performance of  deep learning classifiers trained on them and then test them on the over 105 million unlabeled tweets from March 1 to May 15, 2020 to monitor how the global emotions vary in concerned topics and report other interesting findings. 

\end{itemize}

This is the first  report  of COVID-19 sentiment over 105 million tweets.  The largest analyzed COVID-19 tweet dataset before our work is 1.8M in \cite{xue2020machine}, by an unsupervised way using topic modeling and lexicon features.

\section{Dataset Construction}
\label{sec:dataset}
\subsection{Data Collection}
We used Twint\footnote{https://github.com/twintproject/twint}, an open source Twitter crawler, to collect our tweet dataset where Twint allows users to specify a number of parameters alongside the query, such as tweet language, time period, etc. 
By forming requests with specified parameters, the resulting response was scraped into JSON documents.
We used a unified query across these languages: “covid-19 OR coronavirus OR covid OR corona OR {\small{\AR{كورونا  }    (corona in English)} } ”. We launched 12 instances on 24 cores for downloading daily updates and historical data up to the March 1. 
Data rates varied slightly throughout the period averaging around a little over a million tweets a day. Tweets were saved as JSON documents, and pooled into a shared medium, to be pre-processed and consumed by the language models for sentiment analysis, running on a GPU server (GTX 1080ti GPU and 20 CPUs).

Due to the limited number of tweets in Chinese that could be found on Twitter, our data collection for the sentiment analysis of COVID-19 in China was conducted on Sina Weibo, which is the largest social media platform in China. The Weibo records were collected by Sina weibo API, starting from collecting first the hashtags about COVID-19, and then extracted weibo records including these hashtags.   

\subsection{Data Annotation}
We randomly selected 10K English and 10K Arabic tweets for sentiment annotation. 
These two languages are selected because English and Arabic are among the top-5 popular  languages in the world\footnote{  https://www.vicinitas.io/blog/twitter-social-media-strategy-2018-research-100-million-tweets}. In addition, English can be effectively translated into other languages when needed. We thus have a considerably large labeled dataset for tweet sentimental analysis. 
The sentiment categories were determined by domain experts after reviewing a subset of the collected tweets and discussing for several rounds. The final determined set of labels reflect the complicated sentiments in pandemic. These 10 labels and their covered auxiliary emotions are
\emph{optimistic} (representing \emph{hopeful, proud, trusting}), \emph{thankful} for the efforts to combat the virus, \emph{empathetic} (including \emph{ praying}), \emph{pessimistic} (\emph{hopeless}), \emph{anxious} (\emph{scared, fearful}), \emph{sad}, \emph{annoyed} (\emph{angry}), \emph{denial} towards conspiracy theories, \emph{official} report, and \emph{joking} (\emph{ironical}). 

We recruited over 50 experienced annotators to make every tweet labeled by at least three annotators. 
Example tweets were provided in advance to annotators with suggested categories.
We allowed each tweet to be assigned to multiple labels, which is in line with the convention. For example, the tweet ``Dear Covid19, Will you please retaliate on our behalf. We're hopeless, helpless, restless, speechless.'' has a mixture of \emph{pessimistic} and \emph{sad} emotions.
We allowed the multi-label annotation with our fine-grained sentiments to support the analysis of the complicated emotions in the pandemic. In order to measure the reliability of the sentiment annotations, we conducted a verification study on the annotated tweets where the final results of the labeled tweets is determined by the majority-voting strategy. Our labeled dataset   can also be used for other events with complex emotions, e.g., public opinion analysis and general election analysis. The commonality is that the emotions of these events are multiple aspects.  

For Chinese weibo, we analyzed the COVID-19 posts and annotated 21,173 Weibo in 7 sentimental categories, such as 
\emph{optimistic, thankful, surprised, fearful, sad, angry} and \emph{disgusted}.

\begin{minipage}{\textwidth}
 \begin{minipage}[t]{0.55\textwidth}
  \centering
     \makeatletter\def\@captype{table}\makeatother\caption{Statics of data collected in each language}
     \begin{tabular}{c|c|c|c|c|c|c}\hline
          & En & Es & Ar & Fr & It& Zh\\\hline
       max & 1.7M &652K& 223K & 132K& 63K &23K\\\hline
       min & 272K & 90K& 52K& 28K&15K&0.6K\\\hline 
       mean & 898K &273K &105K &64K&35K&9.6K \\\hline
       total & 68M &21M &8M &4.9M&2.7M&1.1M \\\hline
    \end{tabular}  
    \label{tab:t1}
  \end{minipage}
  \begin{minipage}[t]{0.4\textwidth}
   \centering
        \makeatletter\def\@captype{table}\makeatother\caption{Correlations of normalized volume}
         \begin{tabular}{c|c|c|c|c}   \hline     
          & Es & Fr & It & Ar\\\hline
          En & 0.94 & 0.96 &0.80 &0.83\\\hline
          Es & & 0.96& 0.82&0.84\\\hline
          Fr & & & 0.84&0.86\\\hline
          It & & & &0.79\\\hline
      \end{tabular}
   \end{minipage}
   \label{tab:t2}
\end{minipage}

\begin{figure}[ht]
    \centering
    \includegraphics[width=0.85\textwidth]{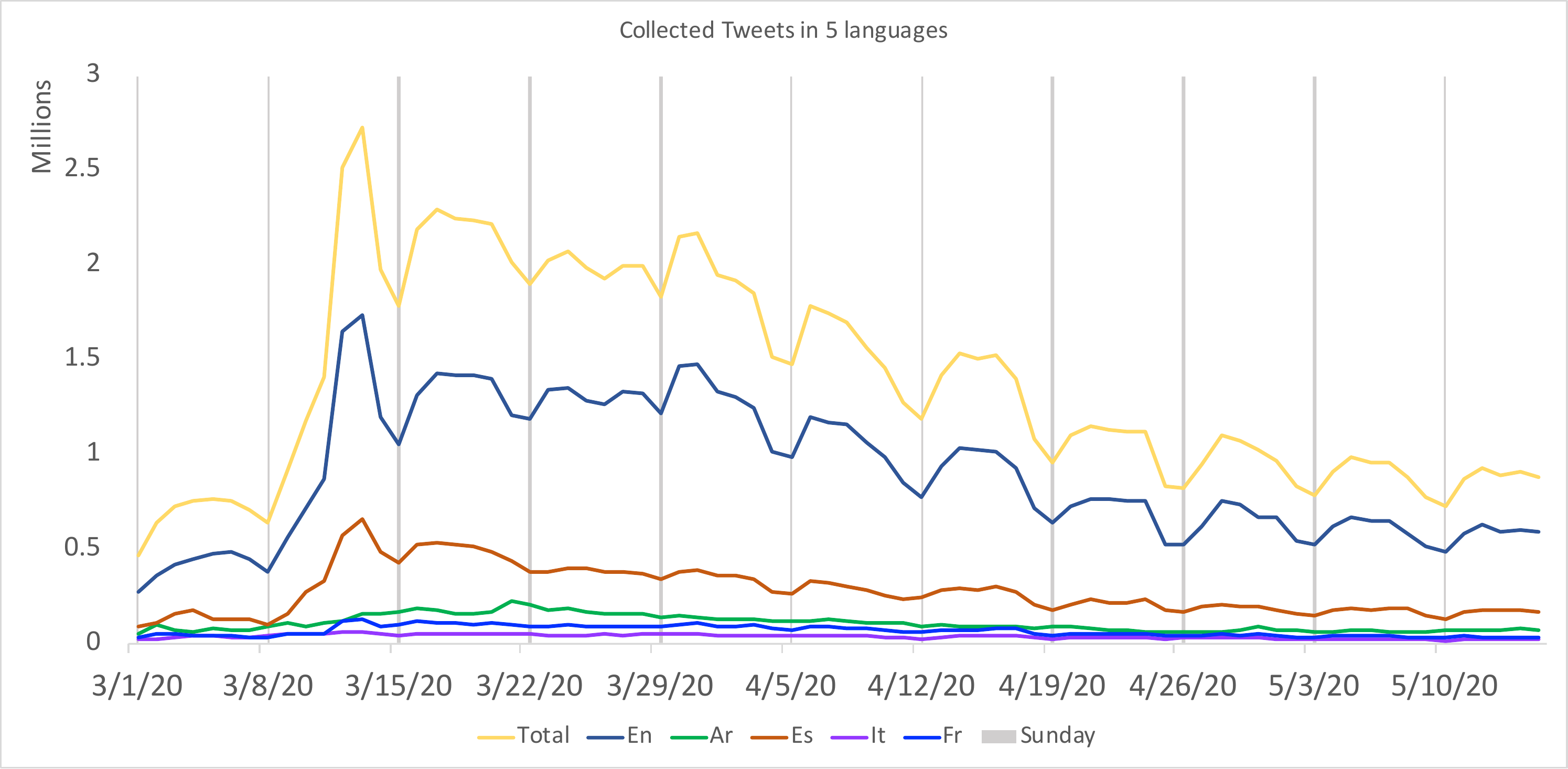}\\
\vspace{+0.3cm}  
    \includegraphics[width=0.85\textwidth]{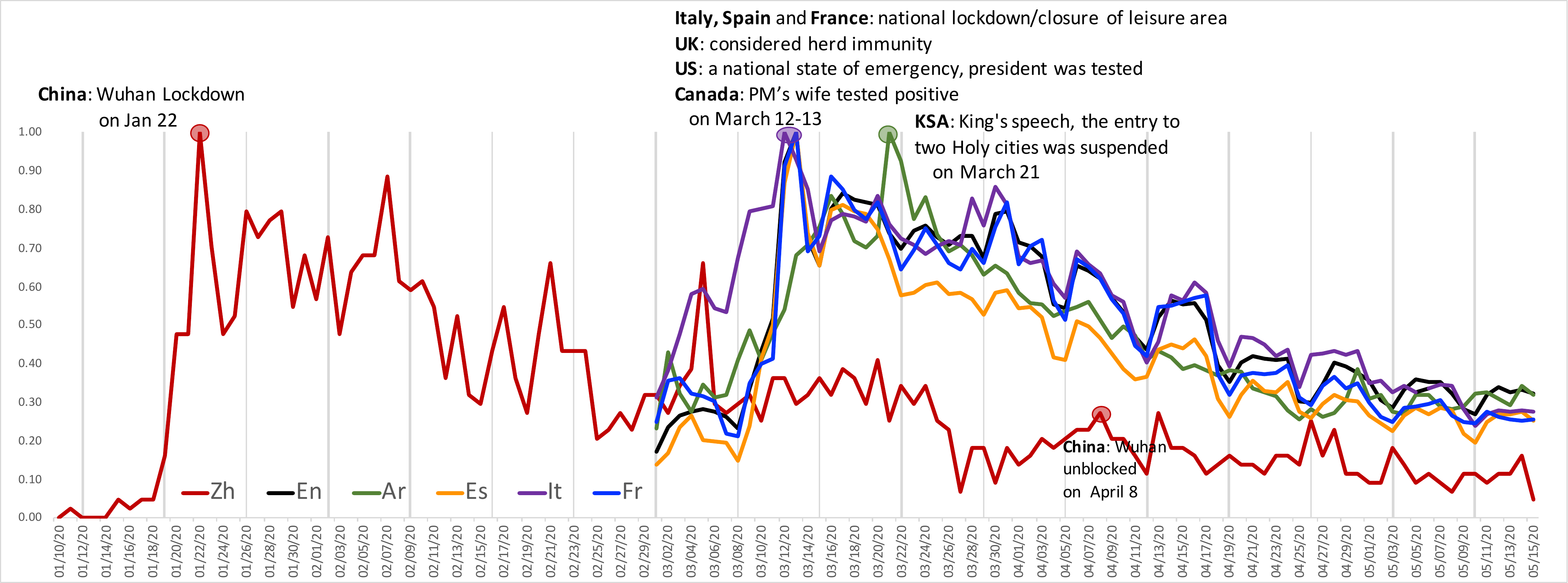}
    \caption{(Top) the absolute dailiy volume of COVID-19 Tweets from Mar 1 to May 15, 2020 collected in 5 languages, English (En), Spanish (Es), Arabic (Ar), French (Fr), and Italian (It). (Bottom) the normalized volume (normalized to maximum number, with volume statistics in Table 1) of the 5 languages,    as well as the  COVID-19 Weibo in Chinese (Zh) from Jan 10 to May 15.} 
    \label{fig:vol}
\end{figure}

\subsection{Data Description}
The processed data is saved and updated in the git repository
using the fetched data. To comply with Twitter’s Terms of Service, we are only publicly releasing the Tweet IDs for unlabeled data and the limited number of Tweet texts for the label data. Hence, the researcher can refetch the original tweet if that tweet is still public.  This dataset is licensed under the Creative Commons Attribution-NonCommercial-ShareAlike 4.0 International Public License (CC BY-NC-SA 4.0). The dataset is available on Github at \url{https://github.com/gitdevqiang/SenWave}.

\subsubsection{Statistics of Unlabeled Tweets}
The volumes of collected daily data for each language are illustrated in Fig.~\ref{fig:vol}. 
The statistics show a similar pattern of rapid rise followed by gradual fall in the global citizen conversation around COVID-19, but with the peak in messages in Chinese reaching the maximum on January 22, two months earlier than the peak in all other languages examined on March 12-13 and March 21, reflecting the lag between the development of the epidemic first detected in Wuhan and the spread to reach pandemic status. 
The discussion and attention quickly reached the peak when important decisions were made. For example, Italy, Spain, and France announced the national lockdown or closure of leisure areas while UK considered the herd immunity and US announced a state of alarm on March 12-13. Meanwhile, in KSA the peak was reached due to the King's speech and the suspended entry to two Holy cities on March 21.
%
%
In addition, people’s attention cools down as time goes on.
As shown in Table 2, the fall and rise in the volume of messages on COVID-19 was remarkably correlated for English, French, Italian, Arabic and Spanish languages, although the rise occurred earlier in Italian, the first western nation to suffer the epidemic. The high correlation coefficient values indicate that populations speaking different languages are responding in a similar way.

\subsubsection{Information of Annotated Tweets}

The label distributions are shown in  Table~\ref{tab:labdist}.  Note that the sum of the percentages is not 1 due to the multi-label annotation in En and Ar. In English, \emph{joking} and \emph{annoyed} emotions took large portions, 
which is consistent with the the reality, since Covid-19 causes deaths,  high unemployment rates and other problems. However, we also see \emph{optimistic} emotion is the third largest category, indicating that people feel confident about  combating the virus and about the future.  
In Arabic, \emph{official} is significantly higher than others, because since the outbreak of Covid-19, most of the Arabic governments announced a lot of decisions regarding different situations on Twitter.
 
\begin{table}[!htp]
    \centering
    \footnotesize
    \caption{The label distributions in annotated English (En) and Arabic (Ar) datasets (\%)} \vspace{-0.2cm}
    \begin{tabular}{c|c|c|c|c|c|c|c|c|c|c}\hline
        & Opti. &  Than. & Empa. & Pess. & Anxi. & Sad &Anno. & Deni. & Offi. &Joki.\\\hline
       En &23.73& 4.98& 3.89& 13.25& 16.95& 21.33& 34.92& 6.31& 12.07& 44.76\\\hline
       Ar &11.27& 3.33& 6.49& 4.65& 7.53& 10.8& 17.17& 2.1& 34.52& 14.18\\\hline
    \end{tabular}
    \label{tab:labdist}
    \begin{tabular}{c|c|c|c|c|c|c|c}\hline
     & Optimistic & Thankful& Surprised& Fearful& Sad & Angry & Disgusted \\ \hline 
  Zh & 16.98 & 14.51 &14.85 &15.24 &13.02 &    15.00 & 10.40  \\ \hline
        \end{tabular}
\end{table} 

\begin{figure}[!htb]
    \centering
    \subfigure[English tweets]{
    \includegraphics[width=0.49\textwidth]{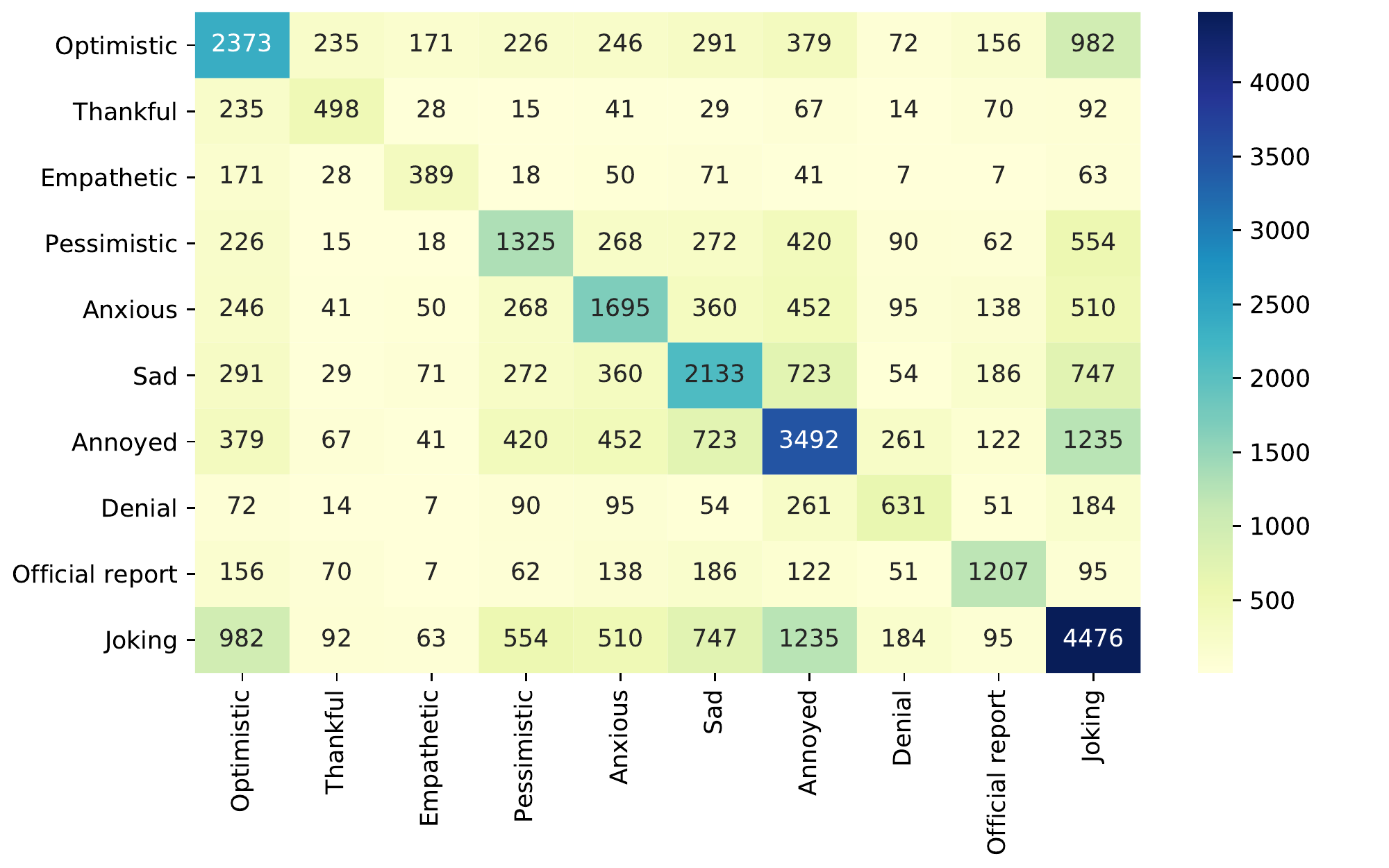}}
    \subfigure[Arabic tweets]{
    \includegraphics[width=0.49\textwidth]{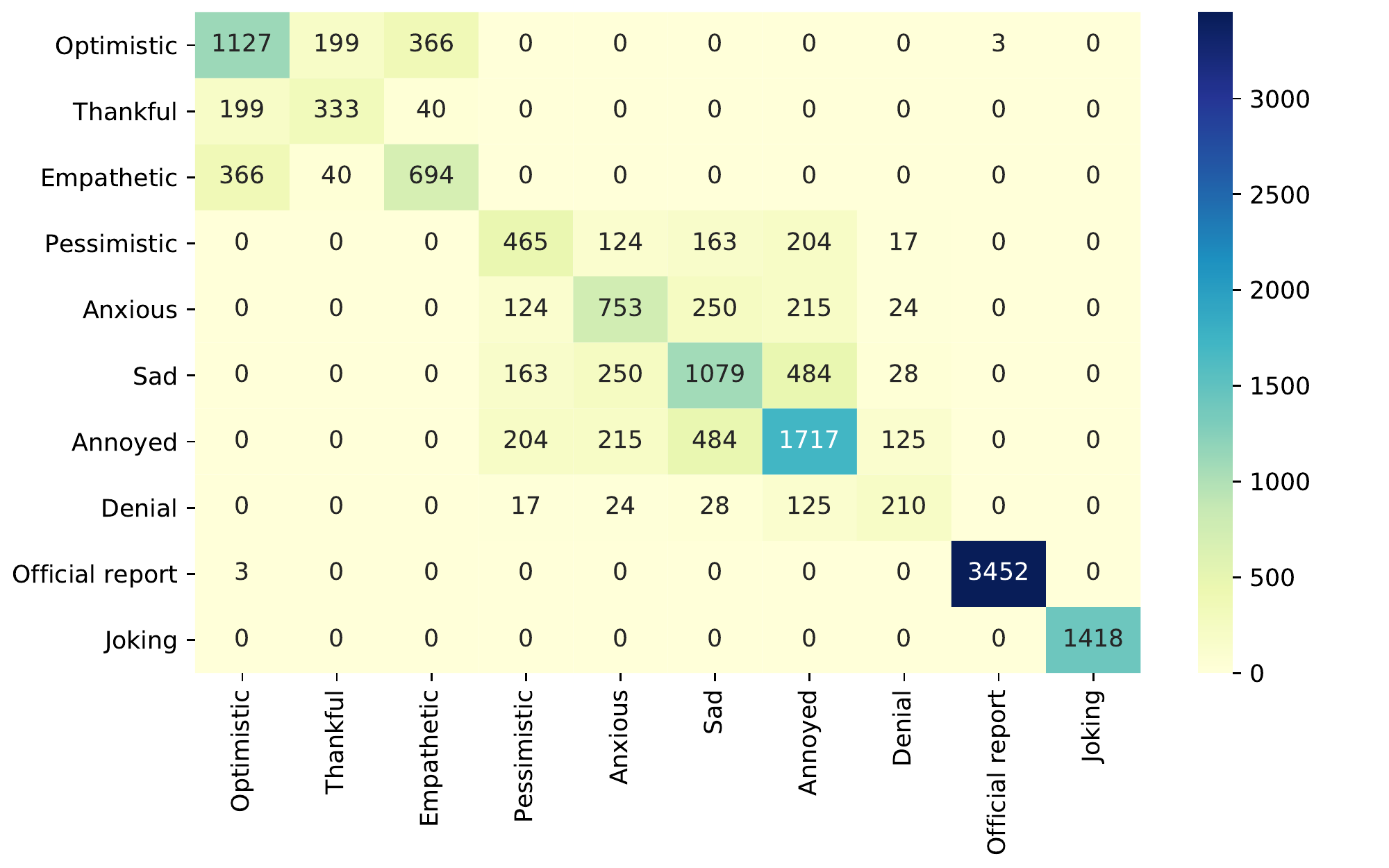}}
    \caption{Heatmaps of co-occurrence of labels for English and Arabic tweets.}
    \label{fig:hm}
\end{figure}

English tweet examples of each category are shown in Table~\ref{tab:examen} where some tweets have more than one label, even with three labels. Based on the statistics of categories, we find that more than 70\% of English tweets were assigned with more than one label, while about 20\% of Arabic tweets were assigned with more than one labels. 
We also present the relations between these labels in Fig.~\ref{fig:hm}. While the label co-occurrence in Arabic shows as three blocks (positive, negative and neural), the  label co-occurrence in English is more complicated. These observations imply that the multi-label classification in our English dataset is more challenging than that in Arabic. We also illustrate this fact in the experiments.

\begin{table}[!htp]
    \centering
    \caption{English examples of each category}
    \scriptsize
    \begin{tabular}{p{2.2cm}|p{10cm}}\hline
      Category   & Example \\\hline
        \multicolumn{2}{c}{Single label} \\\hline
        Optimistic & Nothing last forever, Corona Virus will Vanish this month. “Happy New Month”\\\hline
        Thankful & Gratitude to those who are involved to safeguard our lives from fatal Corona virus. Thanks to them. \#LetUsPrayForCoronaFighters\\\hline
        Empathetic & Allah ap ko bhi safa ata fermain. Ameen. Be strong. IA Allah will give full and speedy recovery. \#coronavirus\\\hline
        Pessimistic &things won't go back to normal \#COVID19 \#coronavirus \#pandemic \#MIT\\\hline
        Anxious & I don’t feel good and I don’t know if I’m just exhausted from working so much or if I have corona\\\hline
        Sad & When someone you know.. apart of your family dies from the Coronavirus it’s shocking; unexplainable. My whole day has been down.\\\hline
        Annoyed & Stop asking to change location man hat how you will spread corona. Fooook\\\hline
        Denial & Unpopular and Insensitive Thought... Corona and Quarantine is a marketing campaign by OTT platforms...!!\\\hline
        Official report & Now schools in Ontario won’t be open until May due to the Coronavirus which might post-pone the 2019-2020 school year to July or August.\\\hline
        Joking & Calling Corona Virus “rona” like she the nastiest little girl in the 5th grade. \#coronavirusmemes \#5G\\\hline
        \multicolumn{2}{c}{Multiple labels} \\\hline
        Empathetic, Sad & So heart breaking any way you see it Prayers to all the families affected by the Covid-19.\\\hline
        Pessimistic, Joking & if i get curved ima go somewhere packed to give myself coronavirus\\\hline
        Anxious, Pessimistic & Does everyone realize we’re going to reach a million cases of this coronavirus by the weekend?\\\hline 
        Denial, Sad, Annoyed & Why is it that no one ever reports on the number of people who recovered from Coronavirus?\\\hline
        Joking, Annoyed & My uncle paranoid about corona virus but still goes to work ....... pick one\\\hline  
    \end{tabular}
    \label{tab:examen}
\end{table}

\section{Model Introduction}
\label{sec:model}
\subsection{Data Preprocessing}
We pre-processed the raw data to ensure the analysis quality. In details, we first remove the  @users, and URLs from the tweet because they do not contribute to the tweet
analysis. Then, we remove emojis and emoticons like $\ddot\smile$ though they can express emotions well since we focused on the analysis textual data. 
Next, we filtered out noisy symbols and texts, which cannot convey meaningful semantic or lexicon information, and may even hinder the model from learning, such as retweet symbol ``RT'' and some special symbols including line break, tabs and redundant blank characters. 
Unlike previous methods which also removed hashtags in tweets, we kept these hashtags since they have meaningful semantics, like ``Proud to be one of the few people who hasn’t texted their ex \#Covid-19 \#Quarantine \#lockdown''. 
Apart from that, we also conducted word tokenization, steaming and tagging with the NLTK tool (https://www.nltk.org/) for  English, Spanish, French and Italian, 
and with Pyarabic for Arabic (https://github.com/linuxscout/pyarabic). We used Jieba for Chinese weibo segmentation (https://github.com/fxsjy/jieba).

\subsection{Multi-label Sentiment Classifiers}
We built our multi-label sentiment classifiers based on deep neural network language models due to their success on diverse NLP tasks. An integration framework called  simpletransformer (https://simpletransformers.ai/) supports the fine-tuning of these pre-trained models and the training of a customized classifier.  We used XLNet~\cite{yang2019xlnet} for English,    AraBert~\cite{antoun2020arabert} for Arabic, and ERNIE \cite{zhang2019ernie} for Chinese (selected due to the better performance of ERNIE than that of Bert \cite{devlin2018bert} and LSTM).

We first evaluated the performance of the sentiment classifier in English and Arabic language on the 10K annotated tweets by 5-fold cross validation, except Chinese on 21K labeled Weibo. Then all 10K labeled tweets were used to train the final sentiment classifiers, except the Chinese sentiment classifier that was trained with 21K Chinese Weibo posts. The trained model was then used for predicting the sentiments of millions of Covid-19 tweets (Mar 1 - May 15, 2020 for Non-Chinese data and January 20 - May 15, 2020 for Chinese Weibo message) for our analysis.

Considering that the translation between English and Spanish, French, Italian has been well developed,   we translated the labeled English tweets into Spanish, French, Italian with Google translate (https://translate.google.com/) to illustrate whether our classifiers can   work well . We manually checked a subset of translated tweets and were surprised by the high quality of translation. We used Bert~\cite{devlin2018bert} for Spanish, French and Italian tweets representation learning, and then follow the same steps for sentiment analysis.

\subsection{Experimental Setting and Evaluation Metrics}
We ran the experiments on a workstation with one GeForce GTX 1080 Ti with memory size 11178MB. The batch size is $16$, and the learning rate is $4e-5$ with $20$ epochs.
We used multi-label accuracy (Jaccard index), F1-macro, and F1-micro, as well as the weak accuracy as the performance metrics. 
The accuracy with Jaccard index is defined as:
\begin{equation}
\nonumber
    Jac.Acc = \frac{1}{|D|}\sum_{i=1}^D{\frac{Y_i\cap \hat{Y}_i}{Y_i\cup \hat{Y}_i}}
\end{equation}
where $Y_i$ is the ground truth labels for the $i$-th testing sample, and $\hat{Y}_i$ is the predicted labels. And the weak accuracy of multi-label classification is defined as:
\begin{equation}
\nonumber
    Acc = \frac{1}{D*m}{\sum_{i=1}^D\sum_{j=1}^ {m} {\sigma(\hat{y}_{ij} == y_{ij})}}
\end{equation}
where $\sigma(\hat{y}_{ij} == y_{ij})$ checks if the predicted $\hat{y}_{ij}$ is the same as the ground truth $ y_{ij}$, which can be 1 meaning the $i$-th testing sample has a $j$-th label, and can be 0 indicating the $i$-th testing sample doesn't have $j$-th label. The total number of corrected prediction of $ y_{ij}$ is averaged with $D$ by $m$, which are the number of  testing samples and the number of labels, respectively.
In addition, we used the  ranking average precision score (LRAP) and Hamming loss,  which are specified for multi-label classification.

\section{Results and Analysis}	
\subsection{Performance of Classifier}
We present the 5-fold cross validation results of our multi-label classifiers on SenWave in Table \ref{tab:perf}. Our classifiers reach above 80\% weak accuracy values, which prove the effectiveness of our models. The multi-label Jaccard accuracy of English and Arabic data is larger/equal than/to $0.5$. However, the accuracy of Spanish, French and Italian tweets are not better than the original data. 
The reasons can be two-folds: 1) the usage of different pre-trained language models: XLNet used for English tweets and AraBert used for Arabic tweets perform better than  Bert generally used for Spanish, French and Italian on the same conditions~\cite{yang2019xlnet,antoun2020arabert}; 2) the difficulty of classifying multi-label English tweets due to the complex multi-labels (shown in Fig.~\ref{fig:hm} (a)). 
We are working on the improvement of these models.  %
It is worth noting that F1 values are around 0.5 due to the class imbalance issue, which will be resolved in our future work. However, the high accuracy, LRAP and low Hamming loss demonstrate that the trained classifiers are usable for practical usage.
The multi-class classification accuracy of Chinese weibo  shown in Table \ref{tab:perf} helped us select ERNIE (with accuracy 0.88) for the final analysis due to its better performance than Bert (with accuracy 0.83) and LSTM (with accuracy 0.78).


\begin{table}[ht]
    \centering
    \footnotesize
    \caption{Validation of the SenWave on the labeled records} 
    \begin{tabular}{c|c|c|c|c|c|c}\hline
          &Acc &Jac.Acc& F1-Macro & F1-Micro & LRAP & Hamming Loss\\\hline
       English &$0.847\pm0.004$ & $0.495\pm0.008$ &$0.517\pm0.012$& $0.573\pm0.008$ & $0.745\pm0.007$& $0.153\pm0.004$ \\\hline
       Arabic &$0.905\pm0.002$& $0.589\pm0.010$ & $0.520\pm0.018$& $0.630\pm0.009$& $0.661\pm0.009$&$0.111\pm0.002$\\\hline 
       Spanish &$0.823\pm0.001$& $0.428\pm0.004$&$0.434\pm0.010$ &$0.511\pm0.003$ &$0.684\pm0.002$&$0.177\pm0.001$ \\\hline
       French & $0.824\pm0.004$&$0.431\pm0.011$& $0.431\pm0.0116$& $0.510\pm0.011$& $0.687\pm0.006$&$0.176\pm0.004$ \\\hline
       Italian & $0.827\pm0.002$&$0.438\pm0.006$ &$0.441\pm0.011$ &$0.516\pm0.005$ &$0.693\pm0.005$&$0.173\pm0.002$ \\\hline \hline
       Chinese  & \multicolumn{6}{l}{$0.88\pm$ 0.001 (ERNIE Acc)} 
       \\\hline
    \end{tabular}    
    \label{tab:perf}
\end{table}

\subsection{Sentiments Variation in Six Languages Over Days}
\begin{figure}[ht]
    \centering
    \subfigure[English]{
    \includegraphics[width=0.49\textwidth]{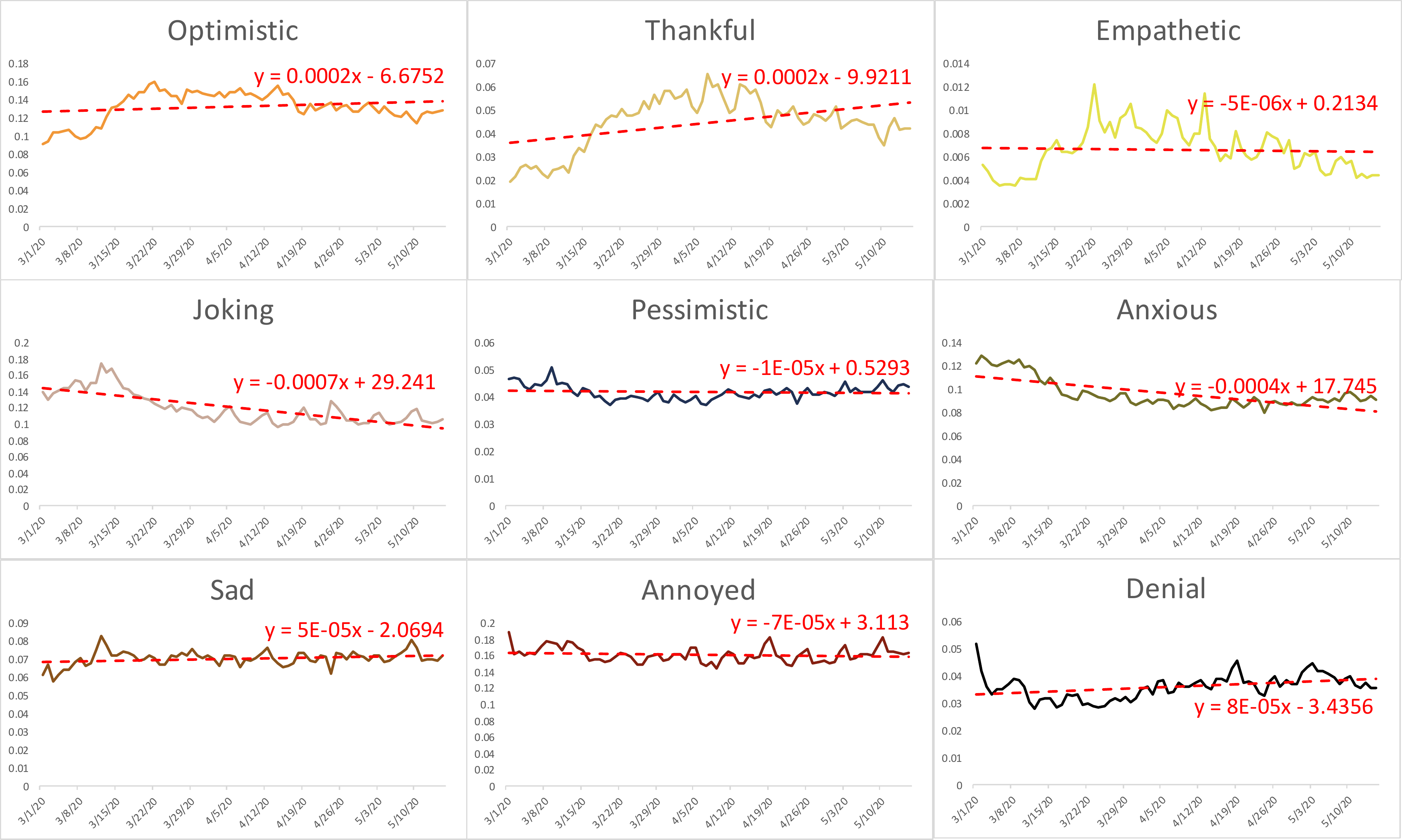}}
    \subfigure[Arabic]{
    \includegraphics[width=0.49\textwidth]{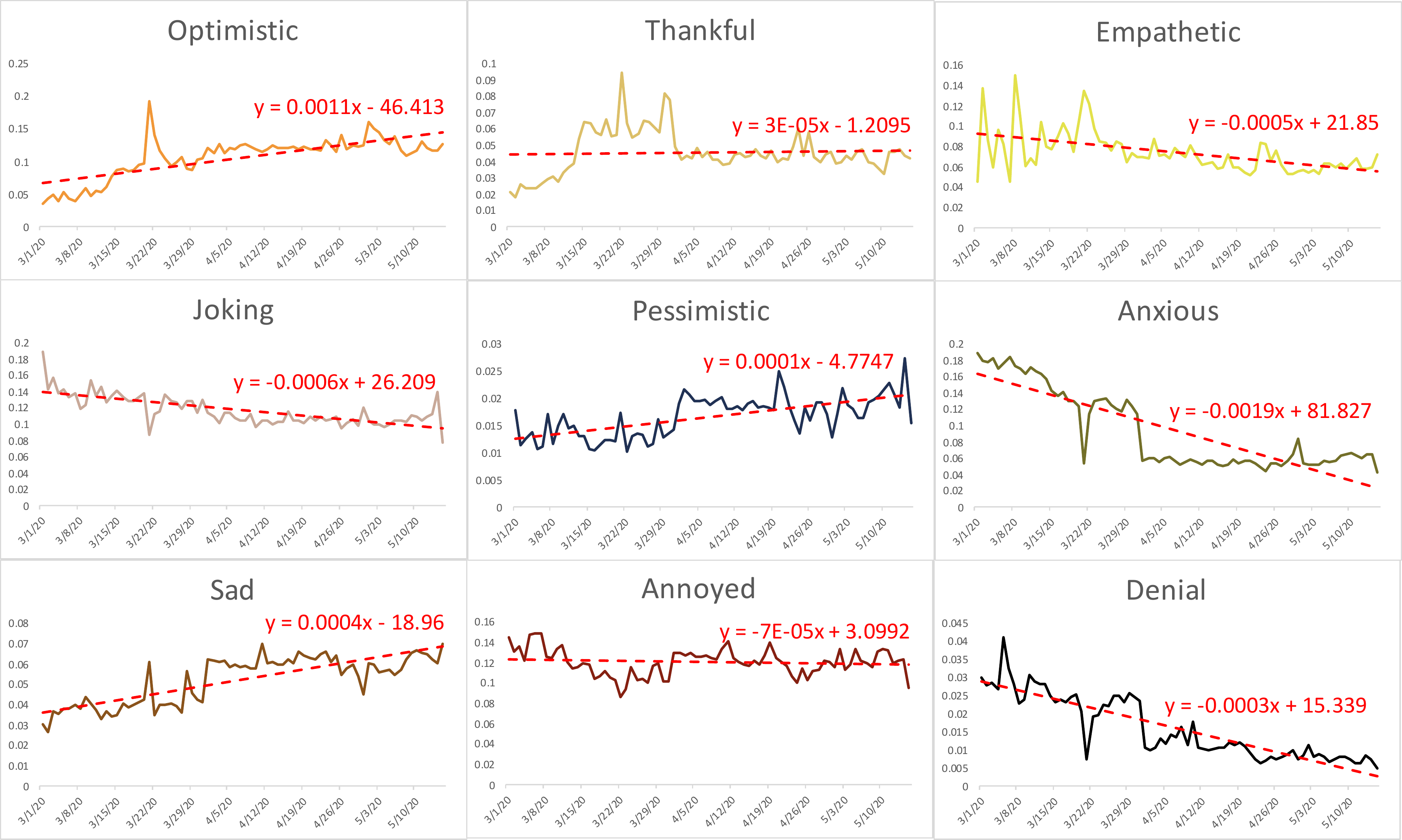}}
    \subfigure[Spanish]{
    \includegraphics[width=0.49\textwidth]{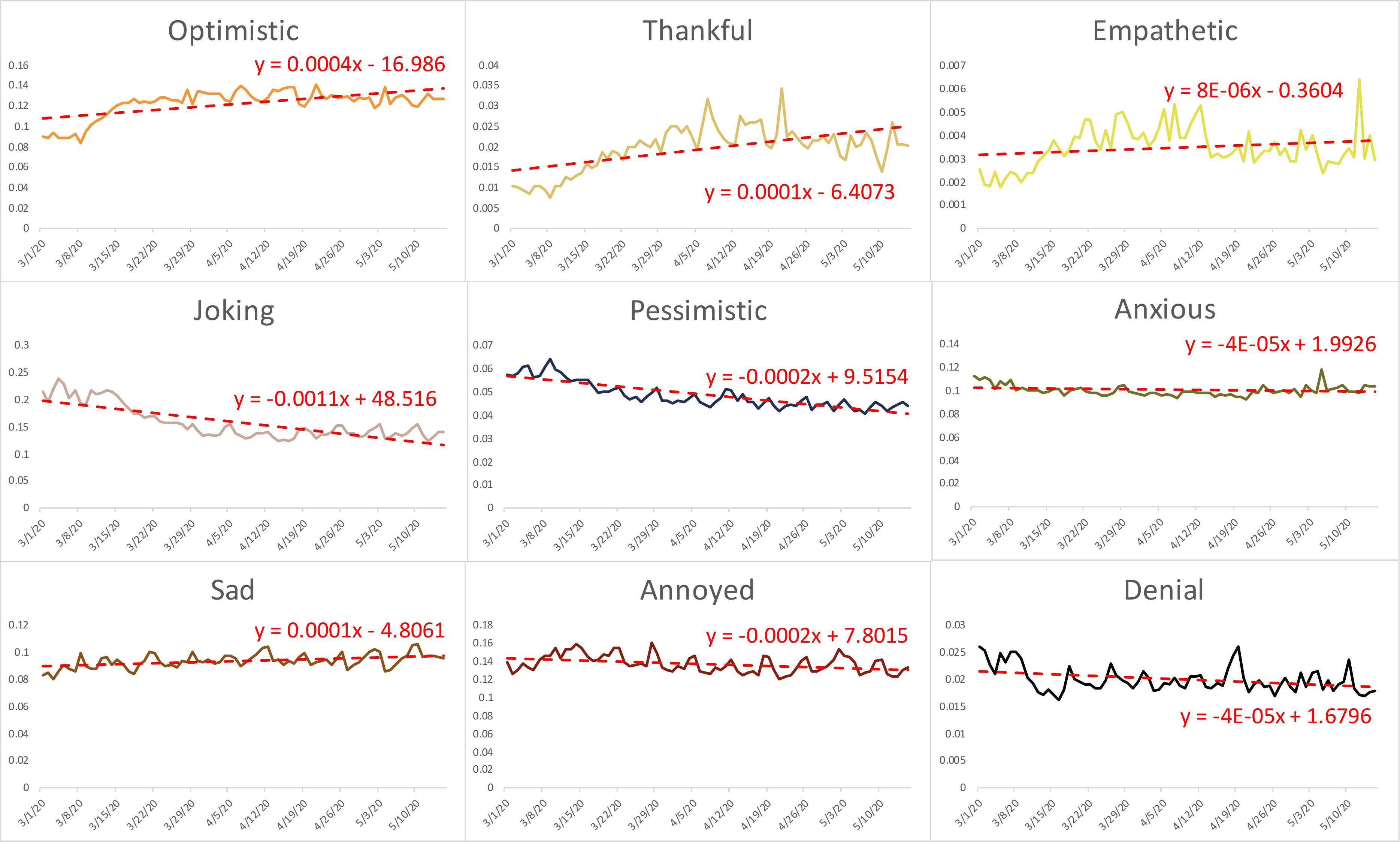}}
    \subfigure[French]{
    \includegraphics[width=0.49\textwidth]{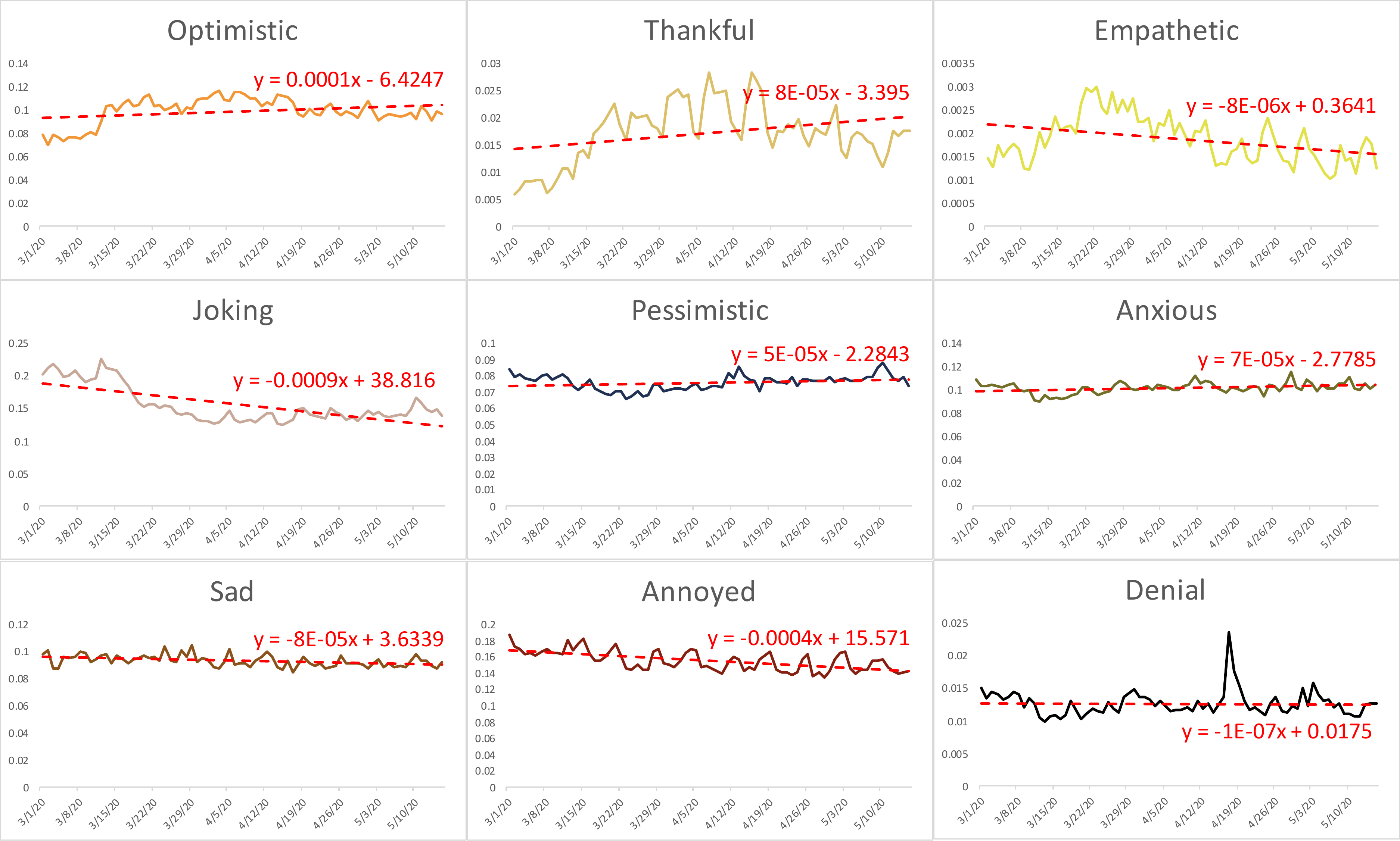}}
    \subfigure[Italian]{
    \includegraphics[width=0.49\textwidth]{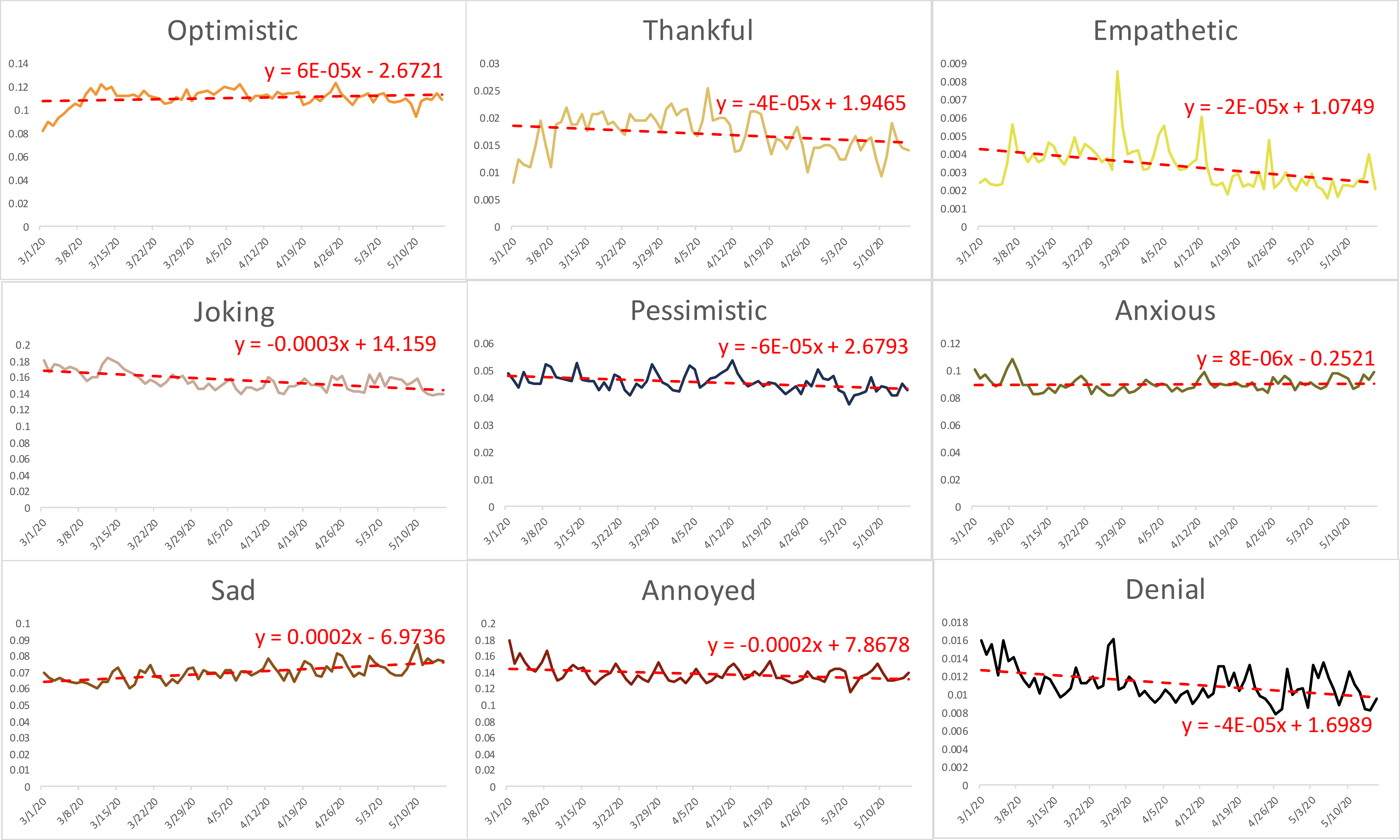}}
    \subfigure[Chinese]{
    \includegraphics[width=0.49\textwidth]{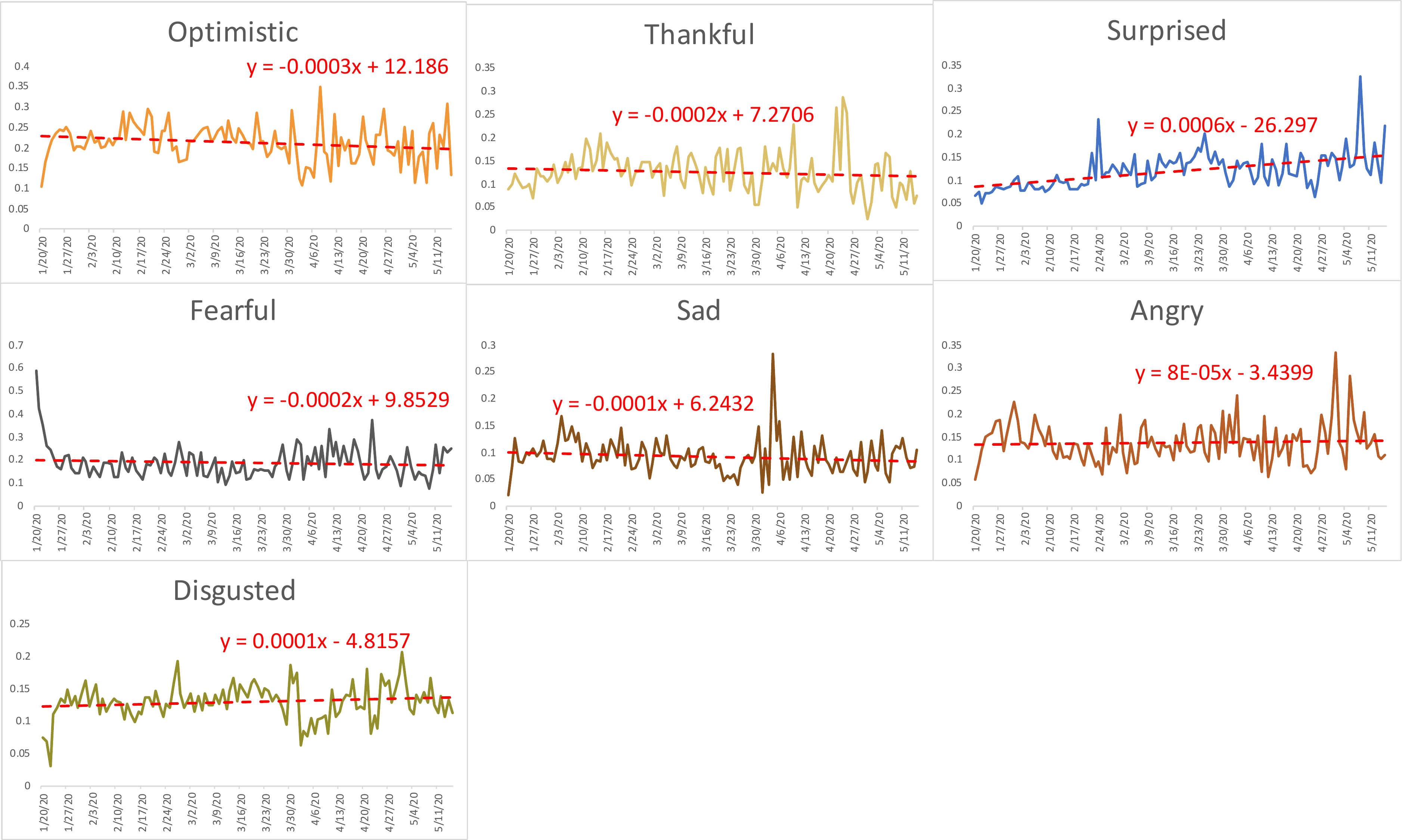}}
    \caption{Sentiment variation of all languages over time. Each subfigure shows the results of one type of languages including 9 emotions for non-Chinese tweets and 7 emotions for Chinese Weibo messages. The linear regression line is fit to each emotion curve, showing the trend of the emotion variation.}
    \label{fig:lan}
\end{figure}

We present the sentiments variation of 6 languages from March 1 to May 15, 2020 for Non-Chinese data and from January 20 to May 15, 2020 for Chinese Weibo message in Fig.~\ref{fig:lan}. The statistics of these sentiment results are given in Table \ref{tab:stat_language}.

The sentiment results of English tweets are  shown in Fig.~\ref{fig:lan} (a). All the positive emotions, including \emph{optimistic, thankful} and \emph{empathetic}, showed the similar trend of first rising up and then falling down. It implies that people first felt positive due to the various decisions  made for combating the virus staring from the mid of March. However, the emotions went down in late April when a large number of people got infected. Among negative emotions,   \emph{anxious} and \emph{joking} fell down with the slope $-0.0004$ and $-0.0007$ respectively as time went on, while the others went stable with slight changes. The \emph{anxious} may be reduced by  the increasing of medical supplies and the fact that people have known much better about Covid-19 than before and got used to the ways living with Covid-19. However, the resulted high unemployment rate and the high number of death may be the reason that \emph{sad} and \emph{annoyed} have been staying high.  

The results of Arabic tweets shown in Fig.~\ref{fig:lan} (b) demonstrate significant variations in all categories of  emotions. 
In particular,   \emph{optimistic} has been rising up, and \emph{anxious, denial} and \emph{joking} are falling down. The \emph{sad} emotion keeps rising up due to the increasing number of new cases in several Arabic-speaking populations, such as Saudi Arabia, Qatar and United Arab Emirates (UAE). 

The rise of \emph{optimistic} and \emph{thankful} and the fall of \emph{pessimistic} and  \emph{annoyed} were also observed in Fig.~\ref{fig:lan} (c) of Spanish tweets. The similar trend of increase in \emph{thankful} is observed in French tweets, as shown in Fig.~\ref{fig:lan} (d). However, the other emotions went stable, except the decline of \emph{joking} and the sudden increase of \emph{denial} to the conspiracy  theory of lab source of corona-virus. 
Italian tweets also showed weak increase or decrease trends in most of the emotions,  as shown in Fig.~\ref{fig:lan} (e), except those in \emph{thankful} and \emph{empathetic}.  

The Chinese Weibo sentiments show strong variations,  but no obvious trend of increase and decrease, as shown in  Fig.~\ref{fig:lan} (f). The most significant decrease in \emph{fearful} is observed in the very beginning on Jan 20, 2020, when   human-to-human transmission was confirmed on that day.
The fearful state continued until January 22, when Wuhan was locked, and the arrival of Chinese New Year.  The significant jumping up of \emph{sad} on April 4 due to the nationwide memorial for victims of Covid-19.

\begin{table}[ht!]
    \centering
        \caption{The statistics of daily sentiment fraction in different categories in all languages, presented as mean$\pm$std, and the number of tweets/weibo from which the statistics were obtained.}
    \scriptsize
    \begin{tabular}{c|c|c|c|c|c|c|c|c|c}\hline
         & Opti. &  Than. & Empa. & Joki.& Pess. & Anxi. & Sad &Anno. & Deni. \\\hline
          En&$0.19\pm0.02$&$0.06\pm0.02$&$0.01\pm0.0$&$0.17\pm0.02$&$0.06\pm0.0$&$0.14\pm0.02$&$0.1\pm0.01$&$0.23\pm0.01$&$0.05\pm0.01$ \\
          &15758721 &5399197 &816664 &13822859 &4721744 &10872331 &8181085 &18445343 &4014464\\\hline
        Es&  $0.18\pm0.02$&$0.03\pm0.01$&$0.0\pm0.0$&$0.22\pm0.04$&$0.07\pm0.01$&$0.14\pm0.01$&$0.13\pm0.01$&$0.2\pm0.01$&$0.03\pm0.0$\\
          & 4284423& 688944&123807 &5472179 &1672356 &3410825 &3203353 &4759697 &671636 \\\hline
         Ar& $0.17\pm0.05$&$0.07\pm0.02$&$0.11\pm0.03$&$0.18\pm0.02$&$0.03\pm0.01$&$0.14\pm0.06$&$0.08\pm0.02$&$0.19\pm0.02$&$0.02\pm0.01$ \\
          &844165 &392832 &627644 &943901 &127120 &794588 &399315 &941615 &138903 \\\hline
         Fr&  $0.14\pm0.02$&$0.02\pm0.01$&$0.0\pm0.0$&$0.22\pm0.03$&$0.11\pm0.01$&$0.14\pm0.01$&$0.13\pm0.01$&$0.22\pm0.01$&$0.02\pm0.0$\\
          & 884553& 158938 &17301 &1351669 &650069 &874491 &814349 &1357922 & 108614\\\hline
         It& $0.17\pm0.01$&$0.03\pm0.01$&$0.01\pm0.0$&$0.24\pm0.01$&$0.07\pm0.0$&$0.14\pm0.01$&$0.11\pm0.01$&$0.22\pm0.01$&$0.02\pm0.0$\\
          & 486093&78086 &15332 &683951 &201141 &391758 &300875 &604932 &49141 \\\hline
          \end{tabular}
       \begin{tabular}{c|c|c|c|c|c|c|c}\hline
         & Optimistic&Thankful & Surprised& Fearful& Sad&Angry&Disgusted    \\\hline
        Zh & $0.21\pm0.05$&$0.12\pm0.04$&$0.12\pm0.04$&$0.19\pm0.07$&$0.09\pm0.03$&$0.14\pm0.04$&$0.13\pm0.03$\\
        &244472 & 143374 &124820 & 206767 & 105974& 155083&143833 \\\hline
    \end{tabular}
    \label{tab:stat_language}
\end{table}


\begin{figure}[ht]
    \centering
    \subfigure[USA]{
    \includegraphics[width=0.49\textwidth]{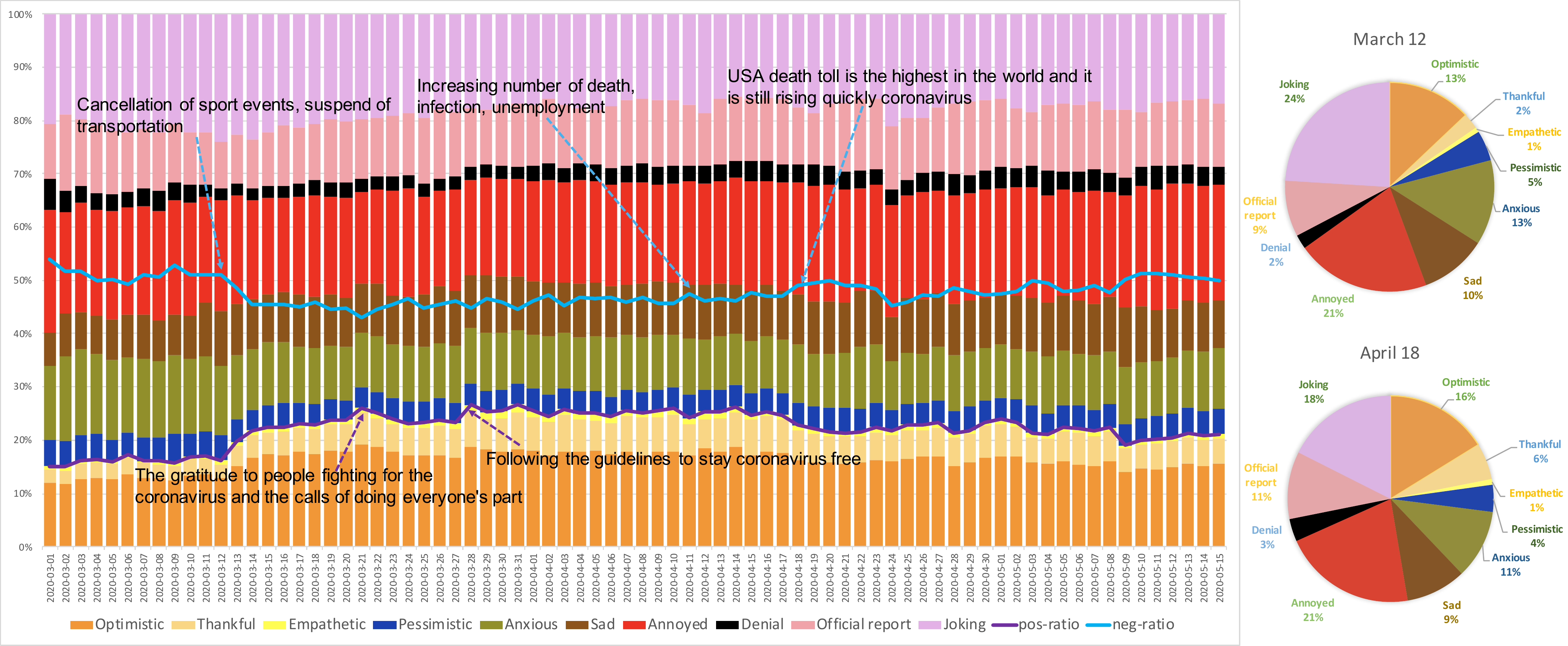}}
    \subfigure[Washington D.C.]{
    \includegraphics[width=0.49\textwidth]{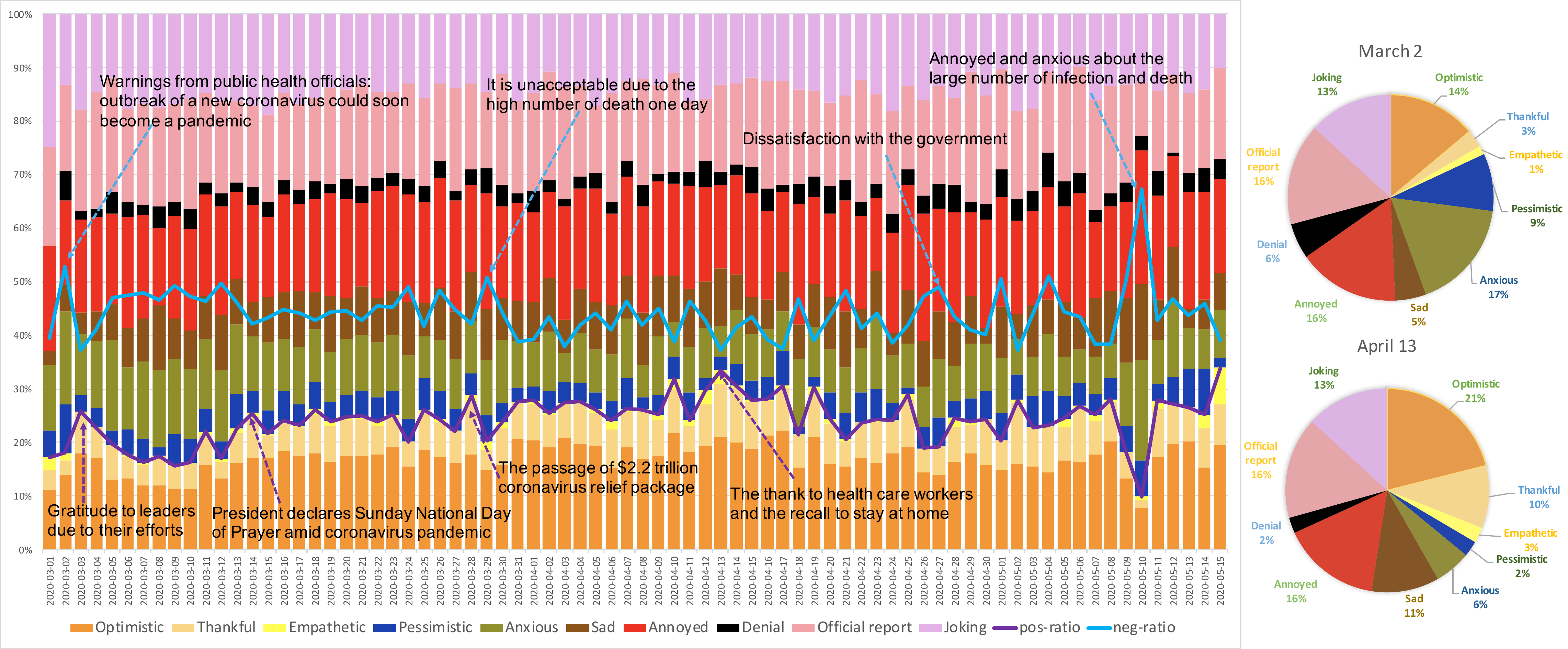}}
    \subfigure[UK]{
    \includegraphics[width=0.49\textwidth]{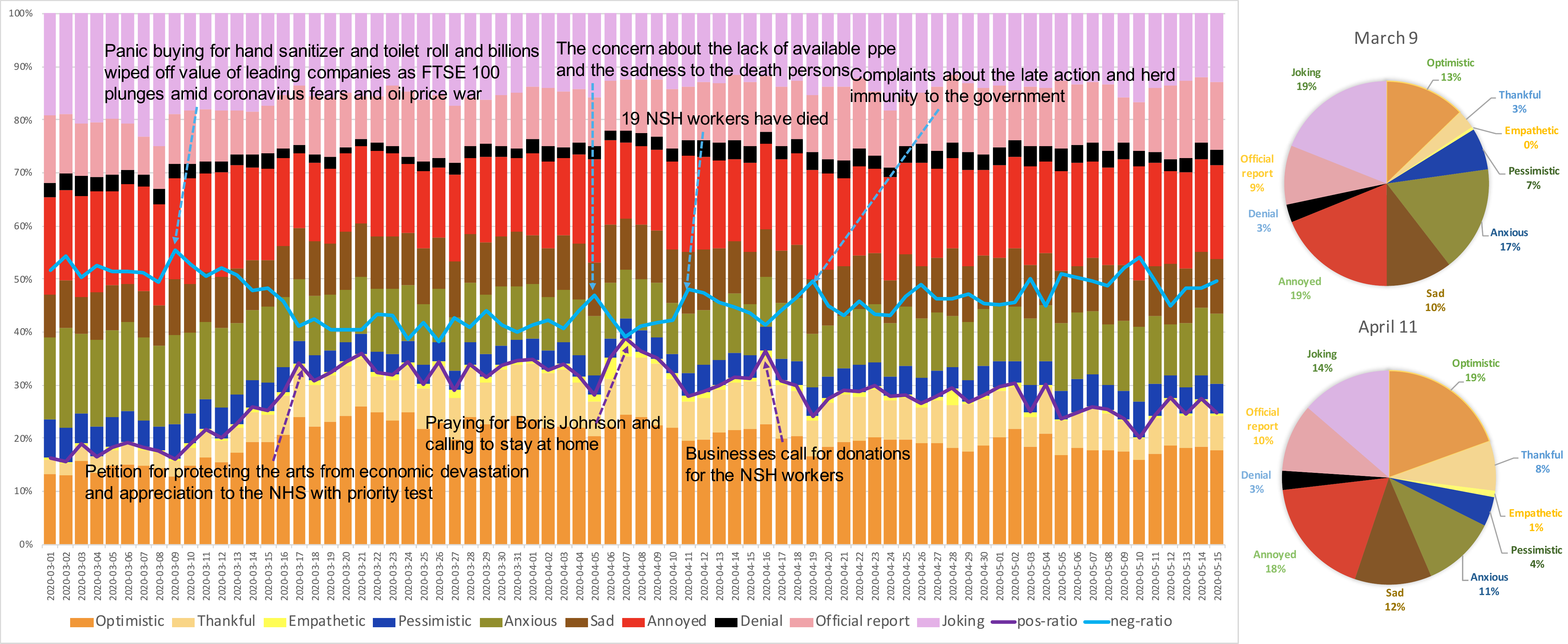}}
    \subfigure[Spain]{
    \includegraphics[width=0.49\textwidth]{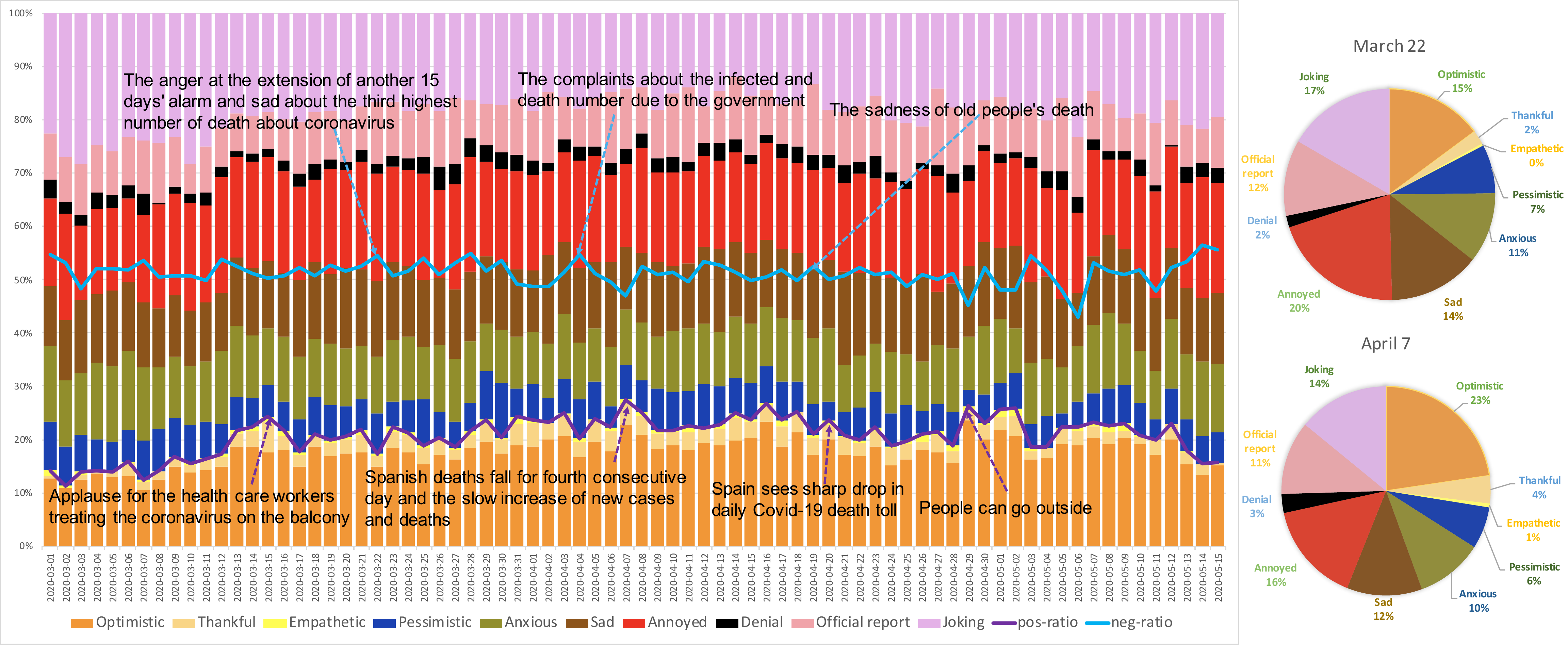}}
    \subfigure[Argentina]{
    \includegraphics[width=0.49\textwidth]{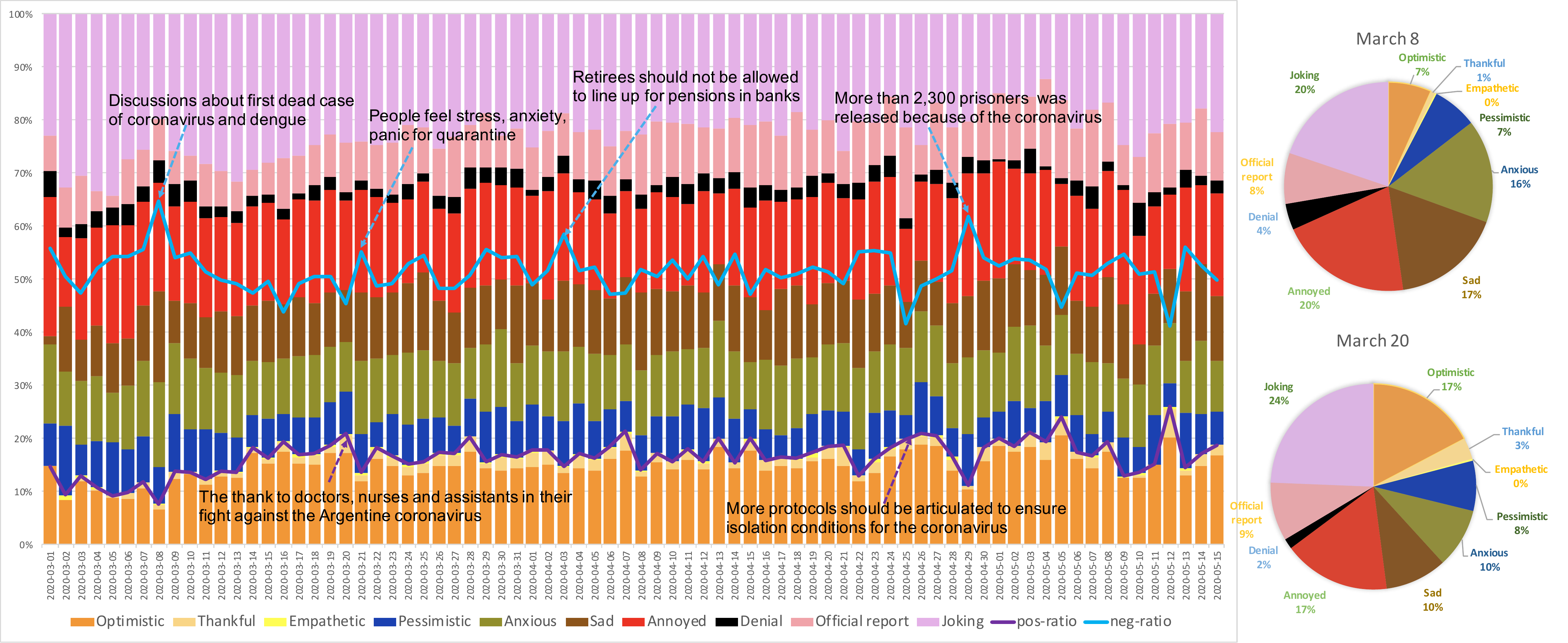}}
    \subfigure[Saudi Arabia]{
    \includegraphics[width=0.49\textwidth]{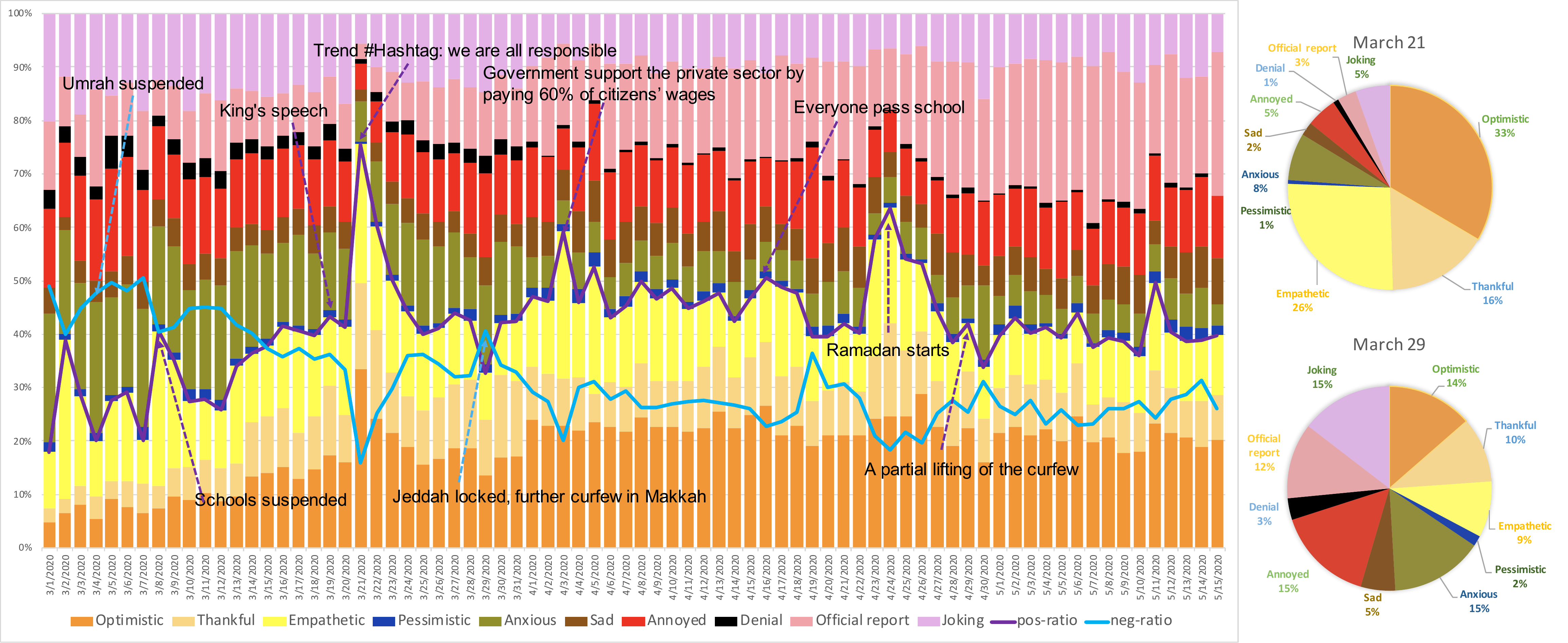}}
    \caption{Sentiment variation in different regions over time. Each bar shows the distribution of sentiments on one day, where sentiments are shown in different colors. The blue curve and purple curve show the positive (sum of \emph{optimistic, thankful, empathetic} in yellow at different intensity) and the negative (sum of \emph{pessimistic, anxious, sad, annoyed, denial} in blue at different intensity), respectively. (Better zoom in to see the interpretation of spikes) }\vspace{-0.1cm}
    \label{fig:areas}
\end{figure}

\begin{table}[ht!]
    \centering
     \caption{The statistics of daily sentiment fraction in different categories of different countries and areas, presented as mean$\pm$std, and the number of tweets from which the statistics were obtained.}
    \scriptsize
    \begin{tabular}
    {c|p{1.23cm}|p{1.23cm}|p{1.23cm}|p{1.23cm}|p{1.23cm}|p{1.23cm}|p{1.23cm}|p{1.23cm}|p{1.23cm}|p{1.23cm}}\hline
         & Opti. &  Than. & Empa. & Pess. & Anxi. & Sad &Anno. & Deni. & Offi. &Joki.  \\\hline
         USA & $0.16\pm0.02$&$0.05\pm0.01$&$0.01\pm0.0$&$0.04\pm0.0$&$0.11\pm0.02$&$0.09\pm0.01$&$0.2\pm0.01$&$0.03\pm0.01$&$0.12\pm0.01$&$0.18\pm0.02$\\
         & 271961&83907 &15011 &67987 &179757 &157241 &318295 &48253 & 184803 &308588\\\hline
        D.C. &$0.17\pm0.03$&$0.07\pm0.02$&$0.01\pm0.01$&$0.04\pm0.01$&$0.1\pm0.03$&$0.08\pm0.02$&$0.18\pm0.02$&$0.03\pm0.01$&$0.17\pm0.03$&$0.14\pm0.02$\\ 
        &2178 &816 &84 &551 &1305 &1064 & 2349&377 &2195 &1846 \\\hline
        UK &$0.2\pm0.03$&$0.07\pm0.02$&$0.01\pm0.01$&$0.05\pm0.01$&$0.12\pm0.02$&$0.1\pm0.01$&$0.17\pm0.02$&$0.03\pm0.01$&$0.11\pm0.02$&$0.15\pm0.03$ \\ 
        &34834 &13220 &1574 &7525 &19233 &16606 &27761 &3977 &17611 &25245 \\\hline
        Spain & $0.18\pm0.03$&$0.03\pm0.01$&$0.01\pm0.0$&$0.06\pm0.01$&$0.11\pm0.01$&$0.13\pm0.02$&$0.18\pm0.02$&$0.02\pm0.01$&$0.09\pm0.02$&$0.19\pm0.04$\\ 
        &4354 &669 &145 &1556 &2807 &3255 &4560 &567 &2246 &4620 \\\hline
        Arge. & $0.15\pm0.03$&$0.02\pm0.01$&$0.0\pm0.0$&$0.07\pm0.02$&$0.12\pm0.02$&$0.11\pm0.02$&$0.18\pm0.02$&$0.03\pm0.01$&$0.09\pm0.03$&$0.23\pm0.04$ \\ 
        &2938 &304 &56 &1452 &2381 &2277 &3654 &504 &1708 &4897\\\hline
        KSA & $0.19\pm0.06$&$0.09\pm0.03$&$0.14\pm0.05$&$0.02\pm0.01$&$0.1\pm0.06$&$0.06\pm0.01$&$0.12\pm0.03$&$0.02\pm0.01$&$0.16\pm0.06$&$0.11\pm0.03$\\ 
        &15137 &7943 &18162 &1269 &8171 &4623 &6597 &385 &13261 &6804 \\\hline  
    \end{tabular}
    \label{tab:stat_area}
\end{table}

\subsection{Sentiments Variation of Selected Areas Over Days}
We selected some countries and areas to illustrate how the sentiments vary over days including USA, Washington D.C. , UK, Spain, Argentina and  Saudi Arabia in Fig.~\ref{fig:areas}. The statistics of these sentiment results are given in Table \ref{tab:stat_area}.

Fig.~\ref{fig:areas} (a) shows in USA, the portion of negative emotions  
is higher than that of positive emotions. 
On March 12, people felt \emph{annoyed} and \emph{anxious} (see the pie charts) since the normal life was affected by coronavirus e.g., cancellation of sport events and suspending of transportation. 
On March 21, however, the positive emotions had a slight increase when people were showing    gratitude for the efforts of healthcare workers.
Similarly, on March 28, people were suggested to follow the guidelines to stay coronavirus free.
However, the negative emotions went up once again due to the increasing rate of death, infection and unemployment on April 11. Several days later on April 18, this figures augmented because USA death toll was the highest in the world and it was still rising quickly.

Fig.~\ref{fig:areas} (b) shows in Washington D.C.,  
on March 2, negative emotions (shown by the blue line) went to the peak due to a  public health official announcement saying that  outbreak of a new coronavirus could soon become a pandemic (see the pie chart at the right hand).  
On March 13, there was a positive spike due to the declaration of a  National Day of Prayer amid coronavirus pandemic.
On March  29, negative pumped up due to  the high number of death within one day achieving 2000, and especially a significant increase of \emph{anxious} because of the negligence of social distancing, even having gathering. 
On April 13, many people showed gratitude to the medical professionals and front-line workers (see the pie chart at the right hand) while during April 27 and 28, people expressed their dissatisfaction with the government due to the resistance, as well as on May 10.
%

Fig.~\ref{fig:areas}(c) shows in UK, on March 9, the negative emotions caused by panic buying of hand sanitizer and toilet roll and people's fear to coronavirus and oil price war leading to the plunging of FTSE 100. 
After different cornoavirus measures were imposed, the  positive sentiment went up significantly. It would be better to zoom in the figures to see other detailed interpretation. 

In Spain (Fig.~\ref{fig:areas}(d)), people applauded for the health care workers treating the coronavirus on the balcony on March 15, felt angry about the extension of another 15 days of alarm and sad about the third highest number of deaths on March 22 (in pie chart). 

In Argentina shown in Fig.~\ref{fig:areas} (e), the proportion of negative emotions was very close to $0.5$ even much higher in some days. On March 8, the discussions about first dead case of coronavirus and dengue were focused on leading to the increase of \emph{anxious}, \emph{sad} and \emph{annoyed} (see pie chart at the right hand).
On March 21, the feelings of stress, anxiety, panic went up because of the long quarantine, which resulted in the increase of \emph{anxious} and \emph{sad}.
On April 29, more than 2,300 prisoners were released because of the coronavirus, which increased the feelings of \emph{pessimistic}, \emph{ anxious} and \emph{annoyed}.

Fig.~\ref{fig:areas} (f) shows stronger positive sentiment in Saudi Arabia than in other countries or areas. Especially, starting from March 13, there was an increase of positive emotions when a lot of decisions were taken by the Saudi government. The peak was reached on March 21, responding to a tweet by the Saudi minister of health: “We are all responsible, staying home is our strongest weapon against the virus”. Another positive peak was shown on April 23-24, when Ramadan started.

\subsection{Sentiments Variation of Studied Topics Over Days}

\begin{figure}[ht!]
    \centering
    \subfigure[Stock market]{
    \includegraphics[width=0.49\textwidth]{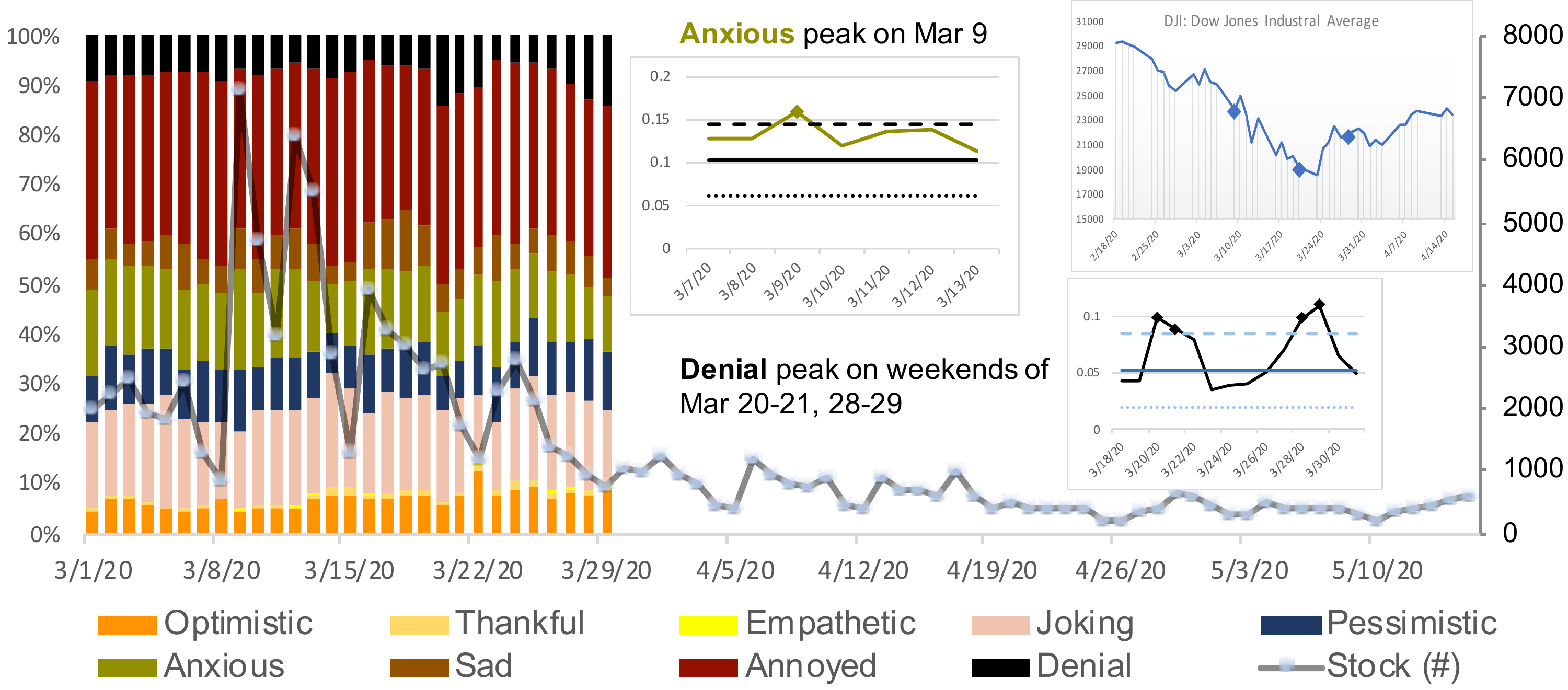}}
    \subfigure[Oil price]{
    \includegraphics[width=0.49\textwidth]{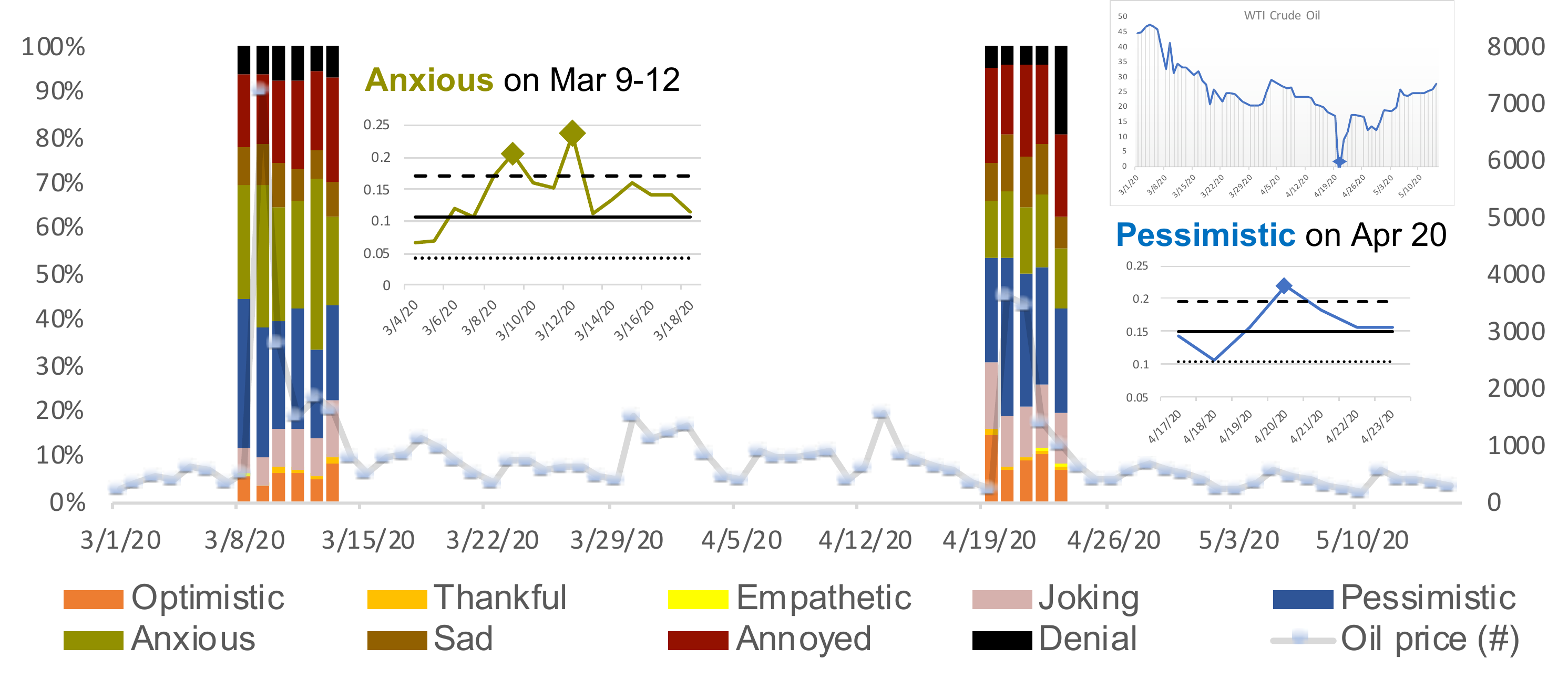}}
    \subfigure[Herd immunity]{
    \includegraphics[width=0.49\textwidth]{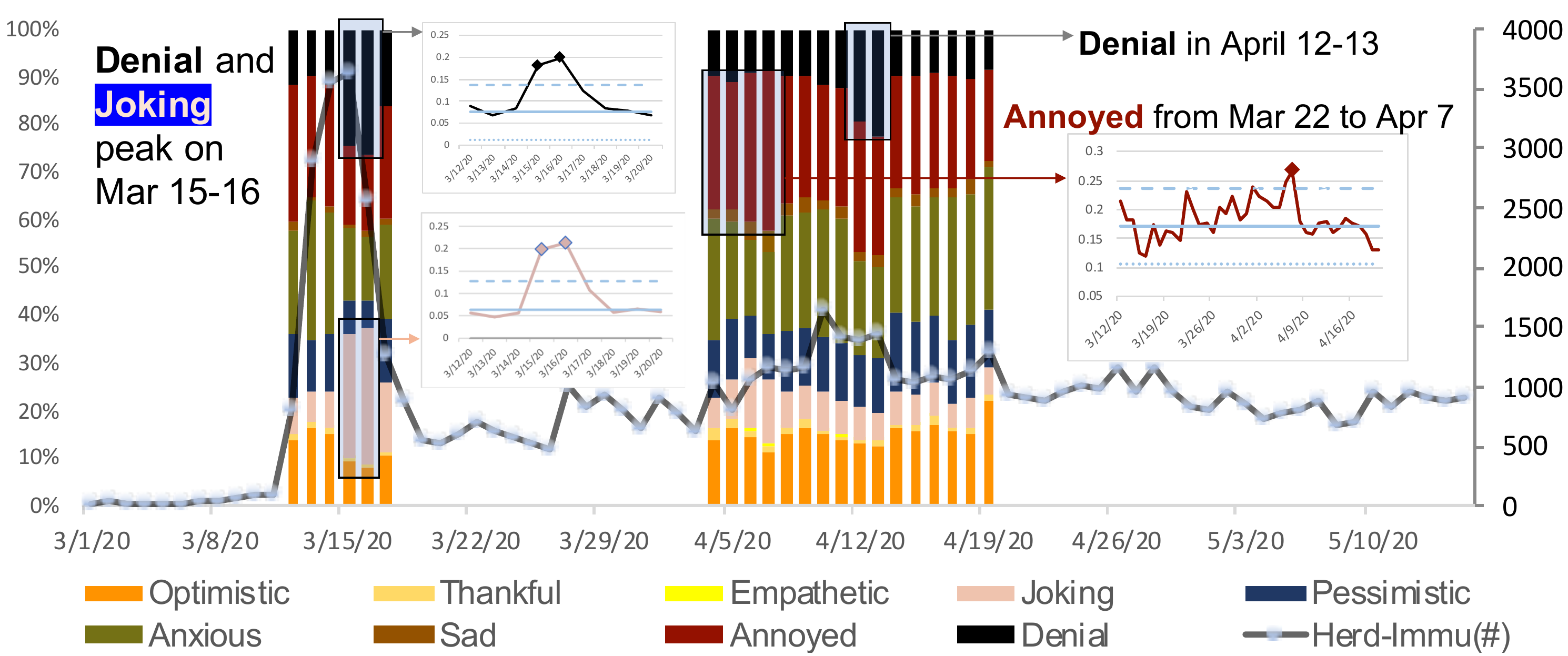}}
    \subfigure[Economic stimulus]{
    \includegraphics[width=0.49\textwidth]{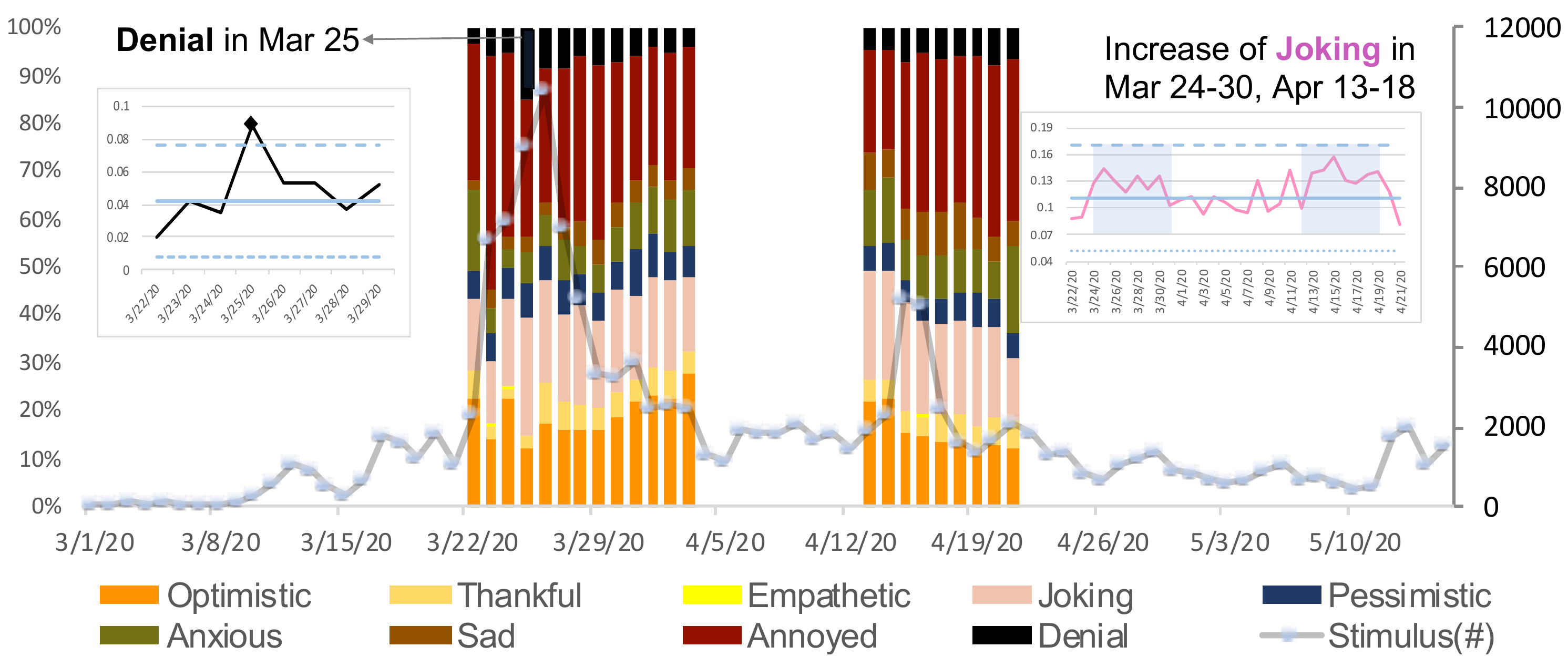}}
    \subfigure[Drug/medicine and vaccine]{
    \includegraphics[width=0.49\textwidth]{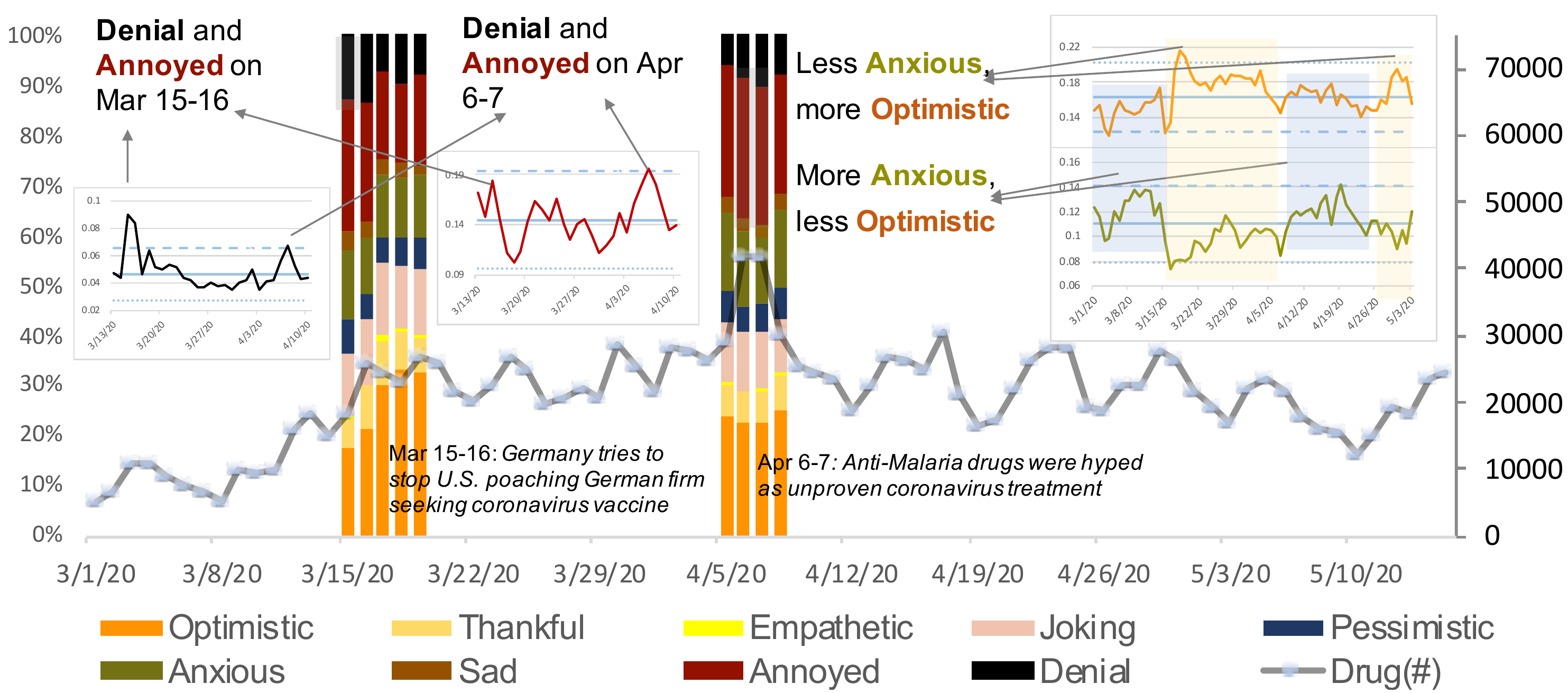}}
    \subfigure[Employment/job]{
    \includegraphics[width=0.49\textwidth]{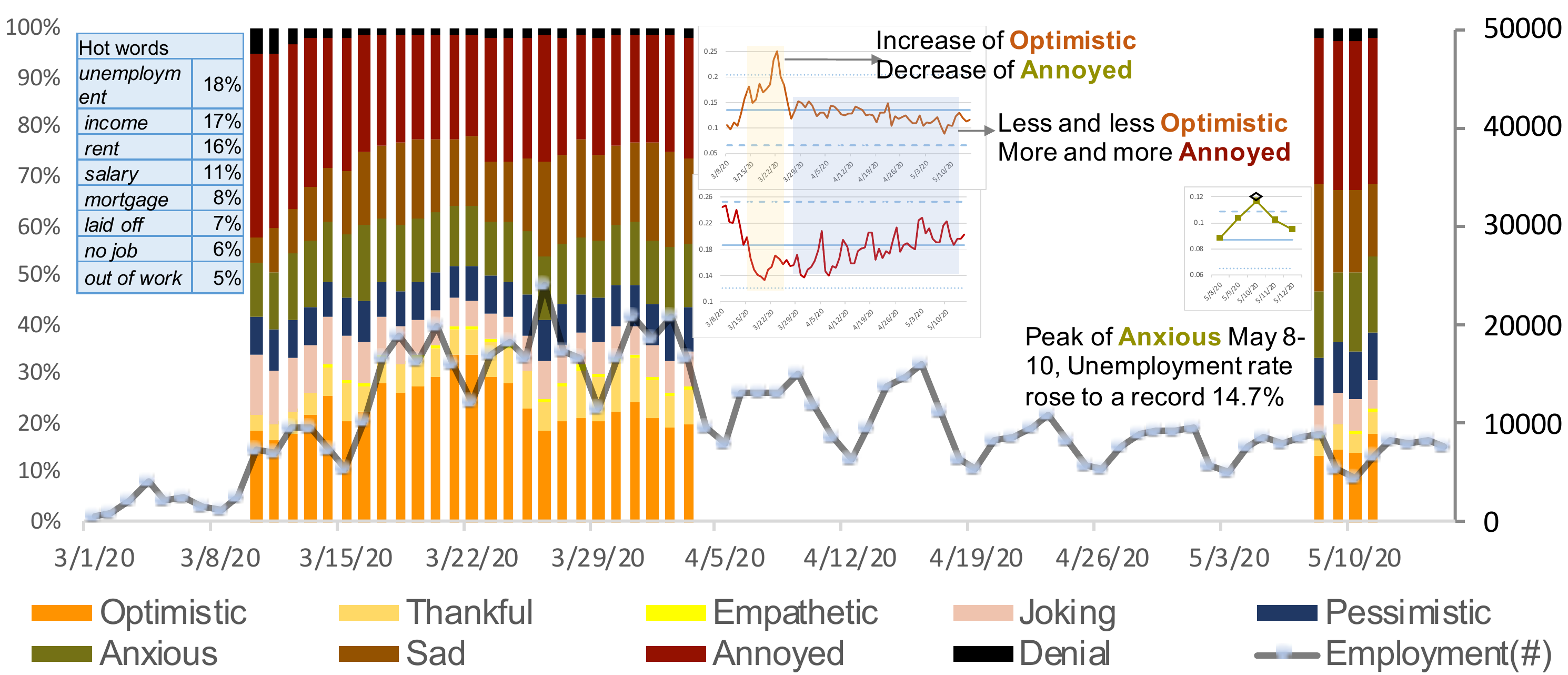}}
    \subfigure[Working from home]{
    \includegraphics[width=0.49\textwidth]{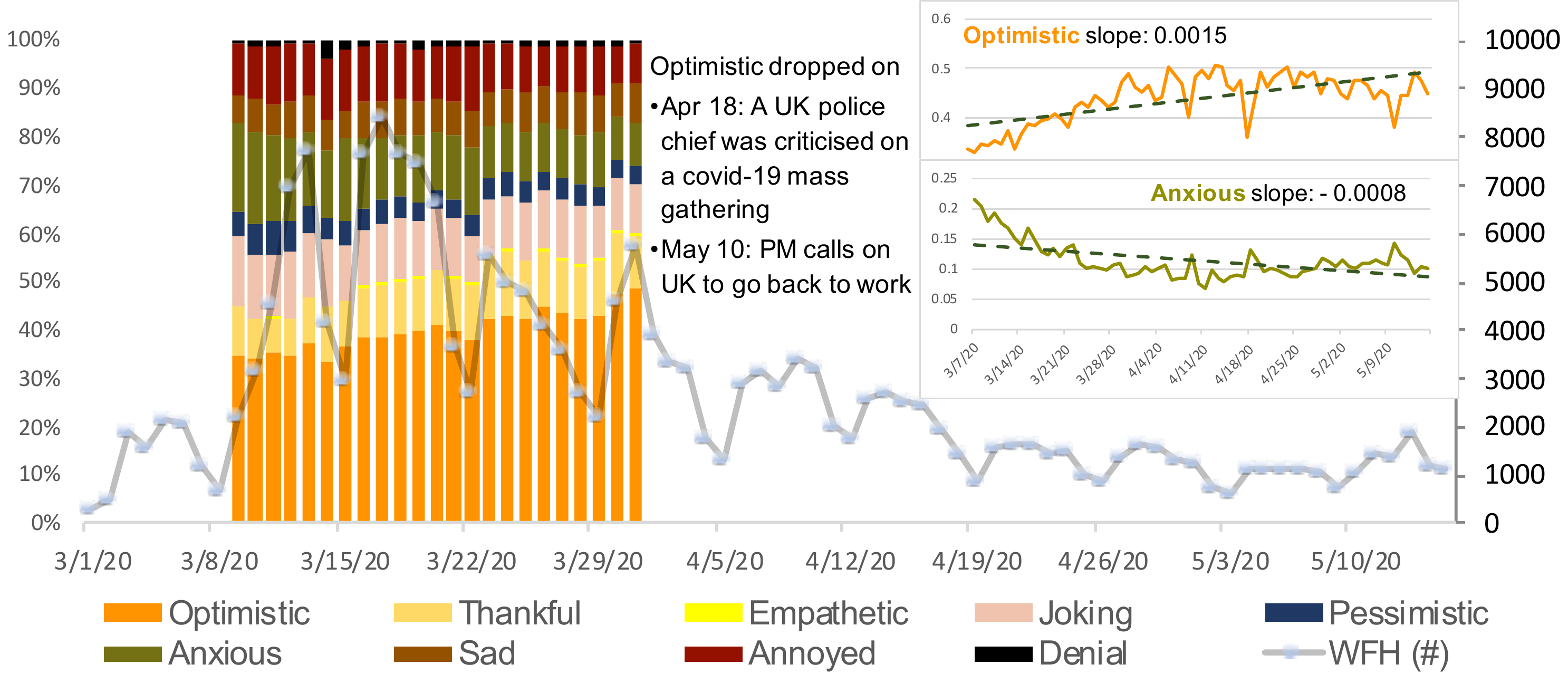}}
    \caption{Sentiments variation on seven topics. We show the sentiment results for these topics when they were intensively discussed (around the peak of volume curve in the background). 
}
    \label{fig:topics}
\end{figure}

We give the sentiment analysis of 7 topics including \emph{stock market}, \emph{oil price}, \emph{herd immunity}, \emph{economic stimulus}, \emph{drug/medicine/vaccine}, \emph{employment/job}, and \emph{working from home}. The results are shown in Figure \ref {fig:topics}.  The statistics of these sentiment results are given in Table \ref{tab:stat_topics}.

The topic stock markets collapsed on March 9,  when the peak of discussion was reached. On this day, \emph{anxious} reaches a high value, which is greater than mean+2*std (out of the black dash line, and the black line is the mean, the dot line is the mean-2*std). On March 12, the DJI (Dow Jones Index) had its worst day since 1987, plunging about 10\% (the second time breakers) and the volumes arrived the second largest. The \emph{anxious} state remained at a high rate during these days.  On the weekends of March 20-21 and March 28-29,  the spikes of \emph{denial} are higher than the blue dash line (mean+2*std), as a reflection of the continuous stock market collapse. DJI: Dow Jones Industrial Average is shown on top of the \emph{denial} curve, and the markers indicate the days of March 9, 20 and 28, when these negative sentiments are reflected on the drop of DJI, as shown in Fig.\ref{fig:topics}(a).

The topic oil price also showed the peak of discussion on March 9. The drop of crude oil price resulted in significant \emph{anxious} on March 9-12. However, this was not the worst. On April 21, crude oil price reached an 18-year low, which is shown on the marked point on the WTI crude oil curve. Among the triggered discussion, we see \emph{pessimistic} was significant. 

The topic herd immunity quickly reached the top on March 14-15 when UK government initially considered it on March 13. Among the intensive discussions from March 13 to 17, \emph{denial} and \emph{joking} were significantly observed on March 15-16. The discussion continued with significant \emph{annoyed} from March 22 to April 7, and caused another rise of \emph{denial} on April 12-13. 

The topic economic stimulus reached top on March 26 when US Senate passed historic \$2tn relief package. And another peak on April 15-16 when the checks were received. Surprisingly, during the discussion in March 23-26, positive was lower compared to other days, and \emph{denial} was significant on March 25. We found many tweets under this topic are for example “this is not enough”, “US economy is tanking”, and “the pandemic is getting worse”. By looking into the \emph{joking}, we see increases in March 24-30 and April 13-18.

The topic drug/medicine/vaccine collected the largest amount of discussion among  these 7 topics (reaching 20-40K on the daily volume). 
This topic has been hot since the global outbreak around March 10. Two events in this topic caused significant \emph{denial} and \emph{annoyed}. The first event was on March 15-16, Germany tries to stop U.S. poaching German firm seeking coronavirus vaccine. The second event was on April 6-7, when Anti-Malaria drugs were hyped as unproven coronavirus treatment. Overall from March to May, we see two sections of more \emph{anxious} and less \emph{optimistic}, and two other sections of less \emph{anxious} and \emph{optimistic}.  

The topic employment/job covered the hot words such as unemployment, income, rent, salary, mortgage, laid off, no job/work etc, as shown in the table included in Fig.\ref{fig:topics}(f). In March, we see the increase of \emph{optimistic} and the decrease of \emph{annoyed}, however in April-May, we see less \emph{optimistic} and the increase of \emph{annoyed}.  Peak of \emph{anxious} was found on May 8-10, when the reported April unemployment rate rose to a record 14.7\% in US.

The topic working from home (WFH) is a warm topic, which obviously has more positive  than all other topics. The bars in Fig.\ref{fig:topics}(g) show that optimistic took more than 40\% in WFH, which is much higher than optimistic in other topics (between 10\% and 25\%) shown in  Fig.\ref{fig:topics}(a-f).  We also see that \emph{optimistic}  keeps increasing with a slope of $0.0015$, while \emph{anxious} is decreasing with a slope of $-0.0008$. Only on two days, \emph{optimistic} dropped: April 18 when a UK police chief was criticised on a covid-19 mass gathering, and May 10 when PM calls on UK to go back to work.

\begin{table}[ht]
    \centering
     \caption{The statistics of daily sentiment fraction in different categories under different topics, presented as mean$\pm$std, and the number of tweets from which the statistics were obtained.}
    \scriptsize
    \begin{tabular}{c|c|c|c|c|c|c|c|c|c}\hline
         & Opti. &  Than. & Empa. (10$^{-2}$) & Joki.& Pess. & Anxi. & Sad &Anno. & Deni.  \\\hline
         Stock& $0.09\pm0.03$&$0.01\pm0.01$&$0.08\pm0.09$&$0.11\pm0.02$&$0.14\pm0.03$&$0.07\pm0.02$&$0.32\pm0.05$&$0.07\pm0.02$&$0.19\pm0.03$ \\
         &10875 &31970 &138 &26890 &15147 &22376 &10325 &48110 &10333\\\hline
         Oil& $0.10\pm0.03$&$0.01\pm0.01$&$0.18\pm0.17$&$0.25\pm0.04$&$0.18\pm0.05$&$0.1\pm0.03$&$0.2\pm0.05$&$0.06\pm0.02$&$0.1\pm0.02$\\
         & 6862&909 &141 &7325 &20122 &15330 &7546 &14755 &4352 \\\hline
         Herd Imm.& $0.18\pm0.05$&$0.01\pm0.01$&$0.19\pm0.55$&$0.13\pm0.03$&$0.24\pm0.05$&$0.02\pm0.01$&$0.23\pm0.04$&$0.1\pm0.04$&$0.09\pm0.04$\\
         & 17698& 1250&108 &11605 &13723 &26275 &2335 & 27006&13865 \\\hline
         Econ. Stim.& $0.19\pm0.05$&$0.05\pm0.02$&$0.13\pm0.25$&$0.07\pm0.03$&$0.09\pm0.03$&$0.05\pm0.03$&$0.3\pm0.05$&$0.07\pm0.03$&$0.18\pm0.04$\\
         &31079 &8717 &169 &33923 &11270 &14986 &8586 &55648 &12483 \\\hline
         Drug&$0.25\pm0.03$&$0.07\pm0.01$&$0.50\pm0.15$&$0.07\pm0.01$&$0.17\pm0.02$&$0.03\pm0.0$&$0.22\pm0.03$&$0.07\pm0.01$&$0.12\pm0.02$ \\
         &537330 &154398 &10749 &249193 &143002 &347756 &70442 &461216 &150892 \\\hline
         Job& $0.2\pm0.05$&$0.06\pm0.02$&$0.41\pm0.18$&$0.09\pm0.01$&$0.13\pm0.01$&$0.15\pm0.04$&$0.28\pm0.04$&$0.02\pm0.01$&$0.07\pm0.02$\\
         & 209210&61106 &4489 &68115 &84671 &127257 &156168 &255922 &20443 \\\hline
         WFH&$0.43\pm0.05$&$0.1\pm0.02$&$0.28\pm0.17$&$0.05\pm0.01$&$0.12\pm0.04$&$0.07\pm0.01$&$0.1\pm0.02$&$0.01\pm0.01$&$0.11\pm0.02$ \\
         & 125176&31970 &975 &32971 &14188 &36042 &21320 &30036 &3271 \\\hline
    \end{tabular}
    \label{tab:stat_topics}
\end{table}

\section{Conclusion}
This paper contributed a Covid-19 sentimental analysis system with annotate datasets, called SenWave, which includes 20K labeled English and Arabic tweets, and 21K labeled Chinese Weibo, as well as 106M+ Covid-19 tweets and Weibo messages collected since Mar 1, 2020 and January 20 respectively.
We trained   classifiers for 6 languages based on deep learning language models to monitor the global sentiments under the Covid-19 pandemic sentimental. We analyzed the sentiments varying of all languages and hot topics over days. On one hand, the emotions on these languages can be directly reflected by corresponding events at the specific days through the varying of volumes and a significant increase of emotions. On the other hand, the sentimental varying trends of 7 topics are also analyzed by showing the corresponding emotions. This work helps to provide a rich resource to the community to study and combat COVID-19.

\bibliographystyle{unsrt}  
\bibliography{references}  

\newpage
\section*{Appendix}
\begin{figure}[ht]
\centering  
\subfigure[Optimistic]{
\label{enwcop}
\includegraphics[width=0.23\textwidth]{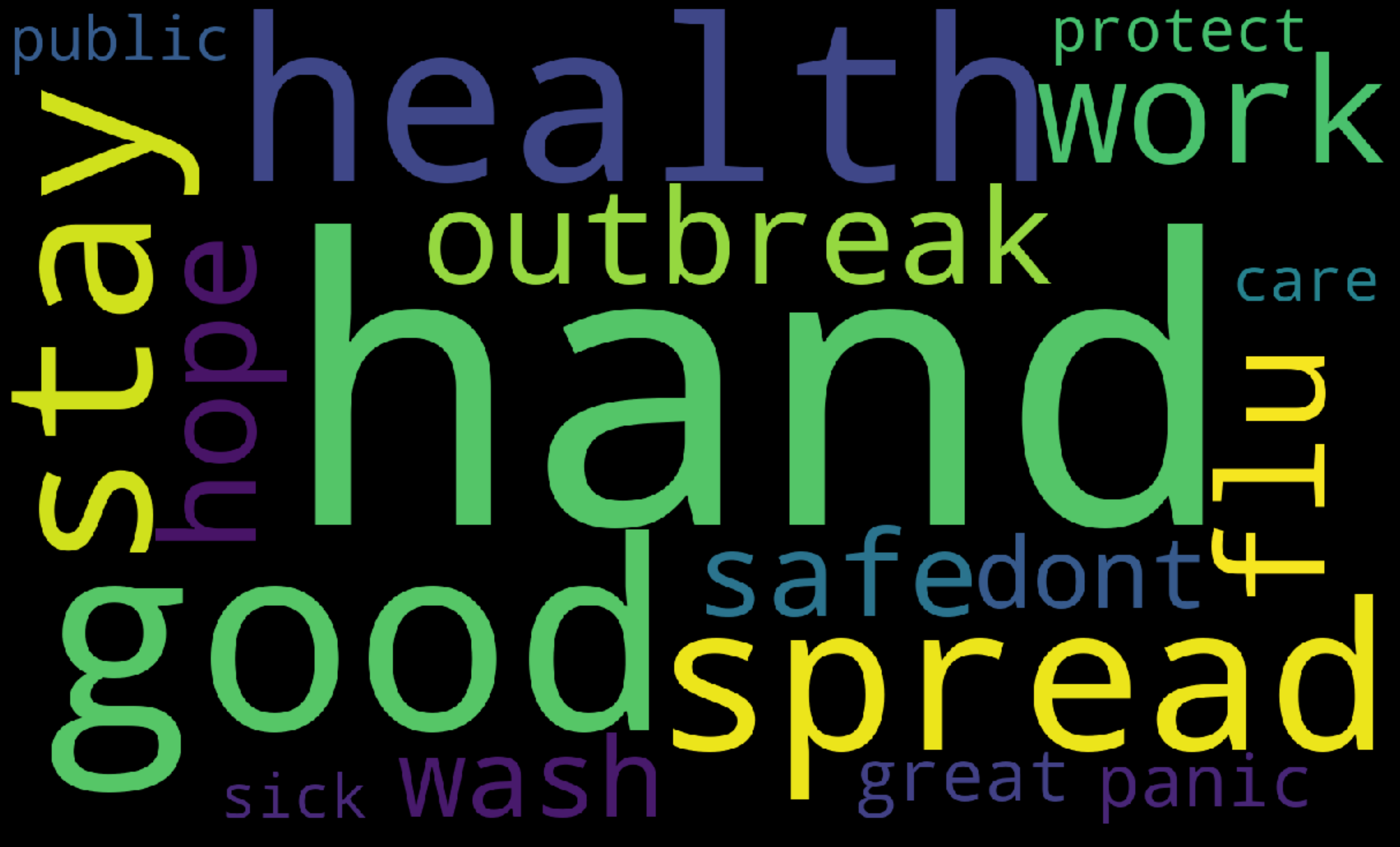}}
\subfigure[Thankful]{
\label{enwcth}
\includegraphics[width=0.23\textwidth]{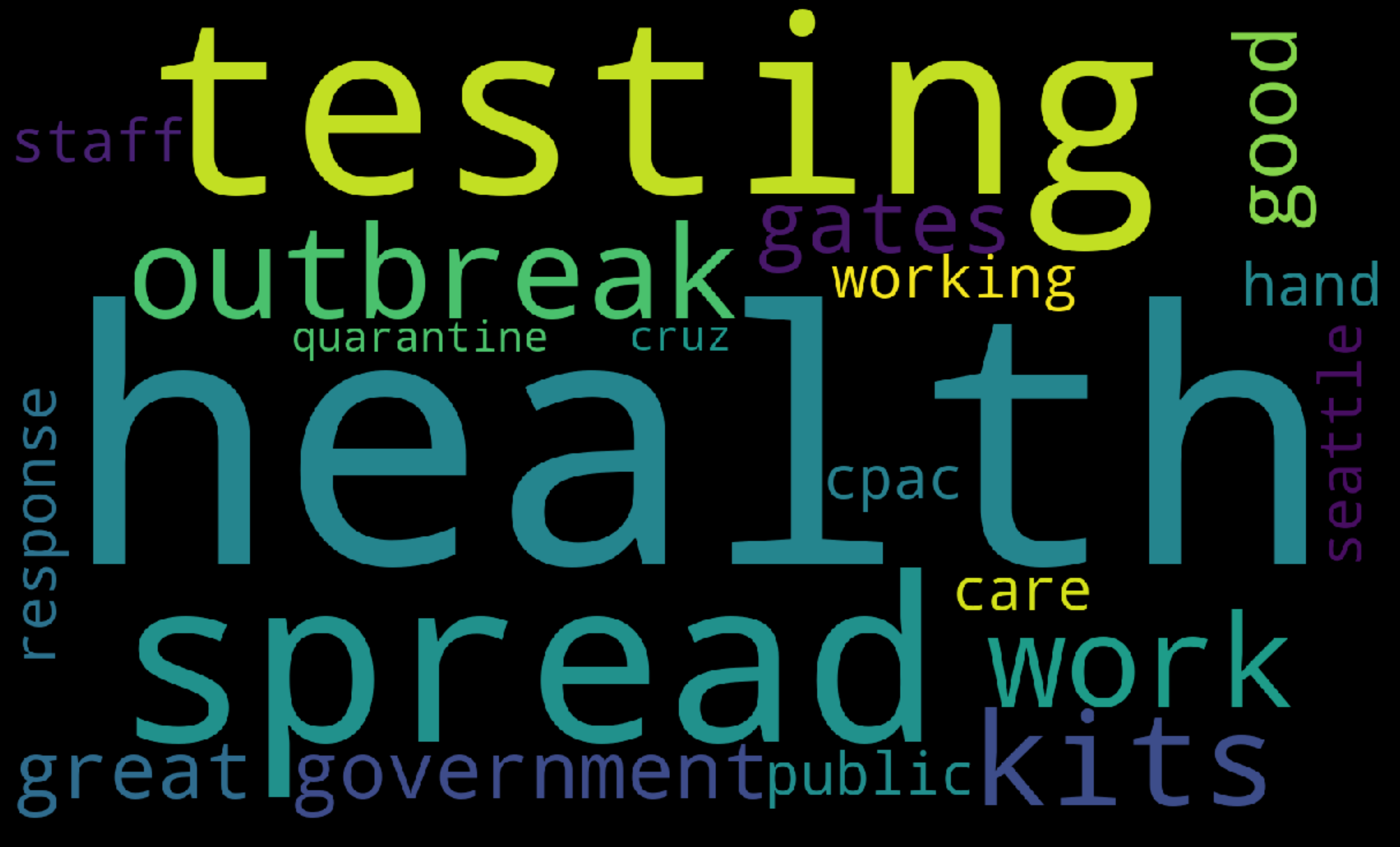}}
\subfigure[Empathetic]{
\label{enwcem}
\includegraphics[width=0.23\textwidth]{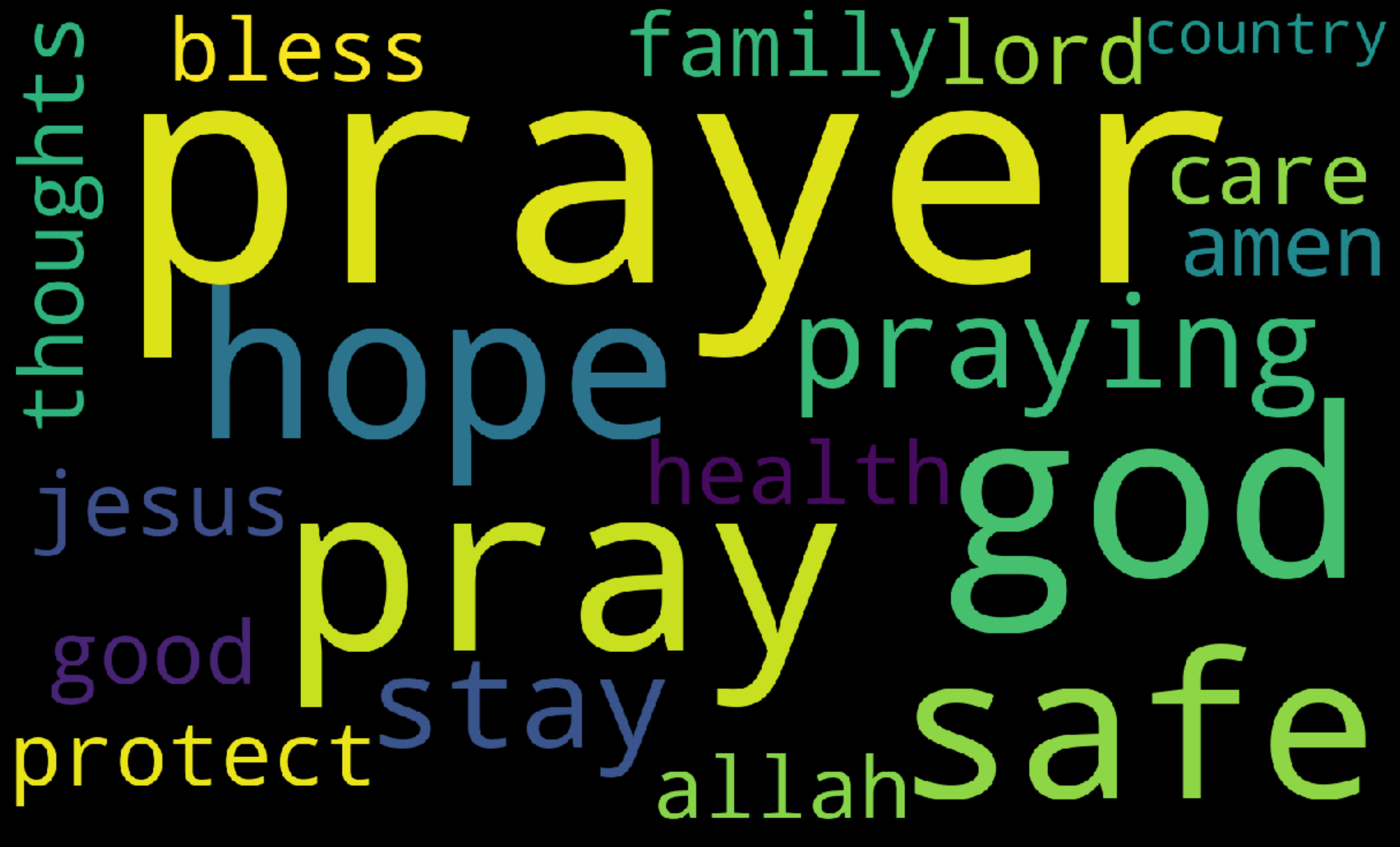}}
\subfigure[Pessimistic]{
\label{enwcpe}
\includegraphics[width=0.23\textwidth]{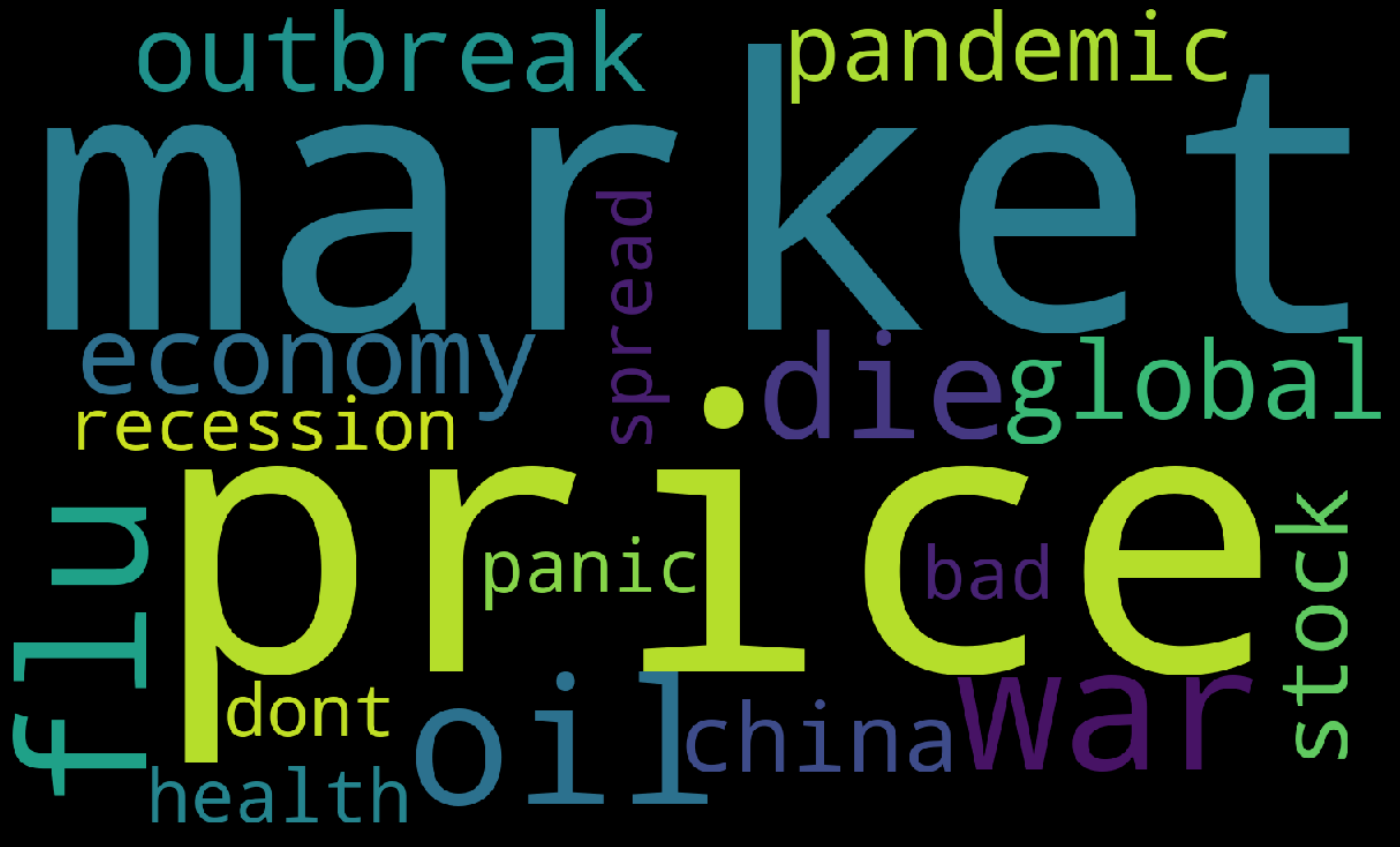}}
\subfigure[Anxious]{
\label{enwcas}
\includegraphics[width=0.23\textwidth]{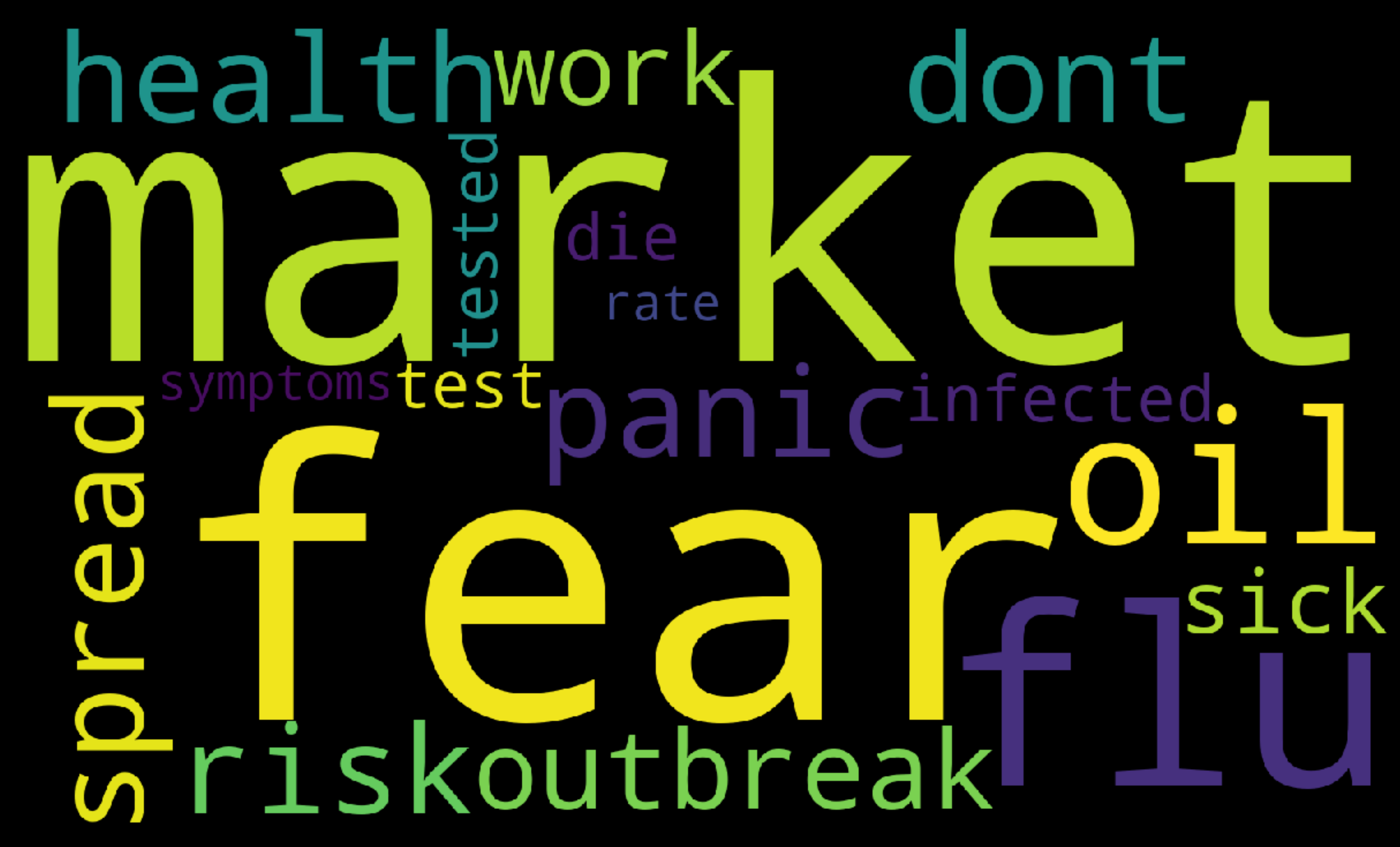}}
\subfigure[Sad]{
\label{enwcsa}
\includegraphics[width=0.23\textwidth]{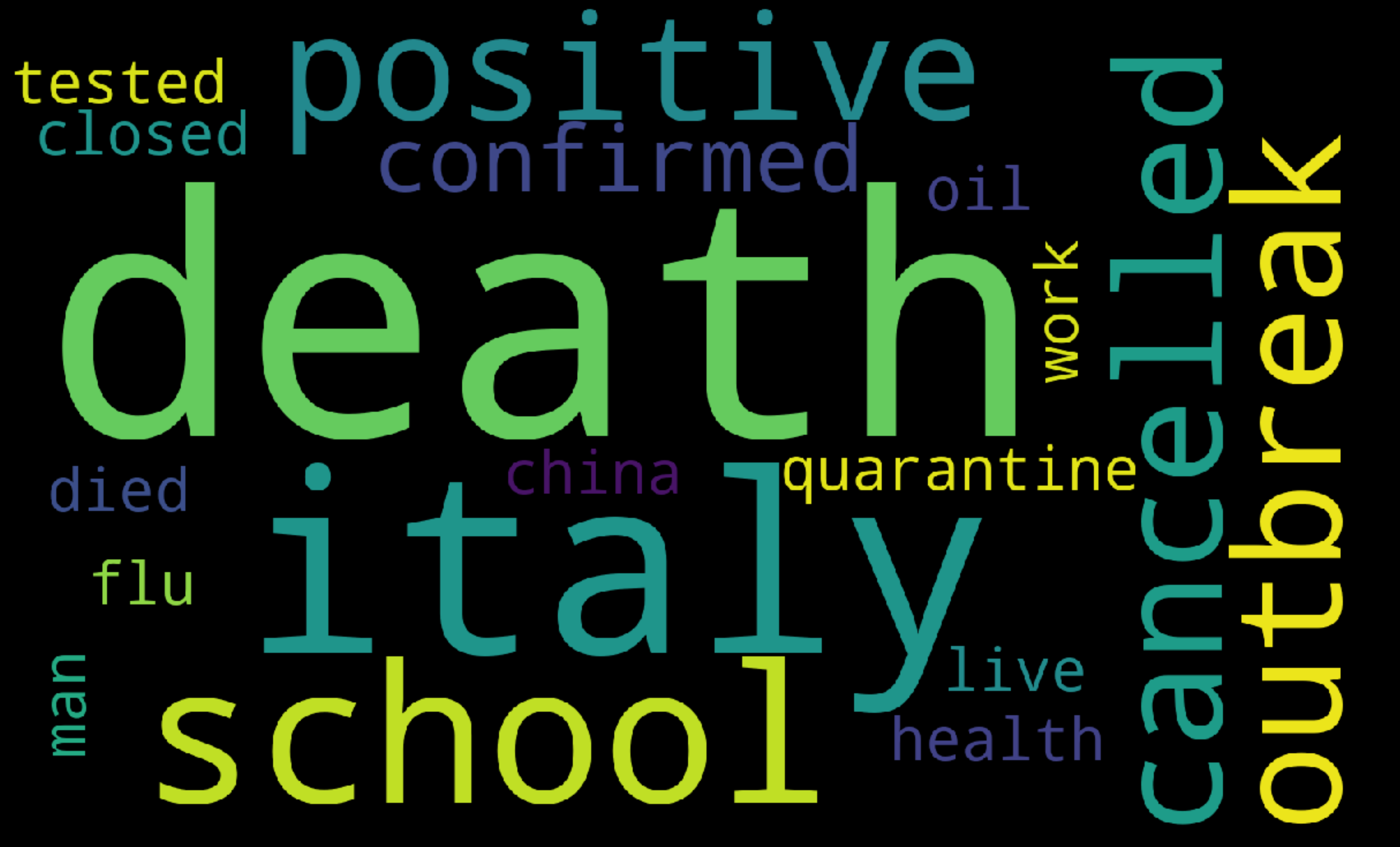}}
\subfigure[Annoyed]{
\label{enwcad}
\includegraphics[width=0.23\textwidth]{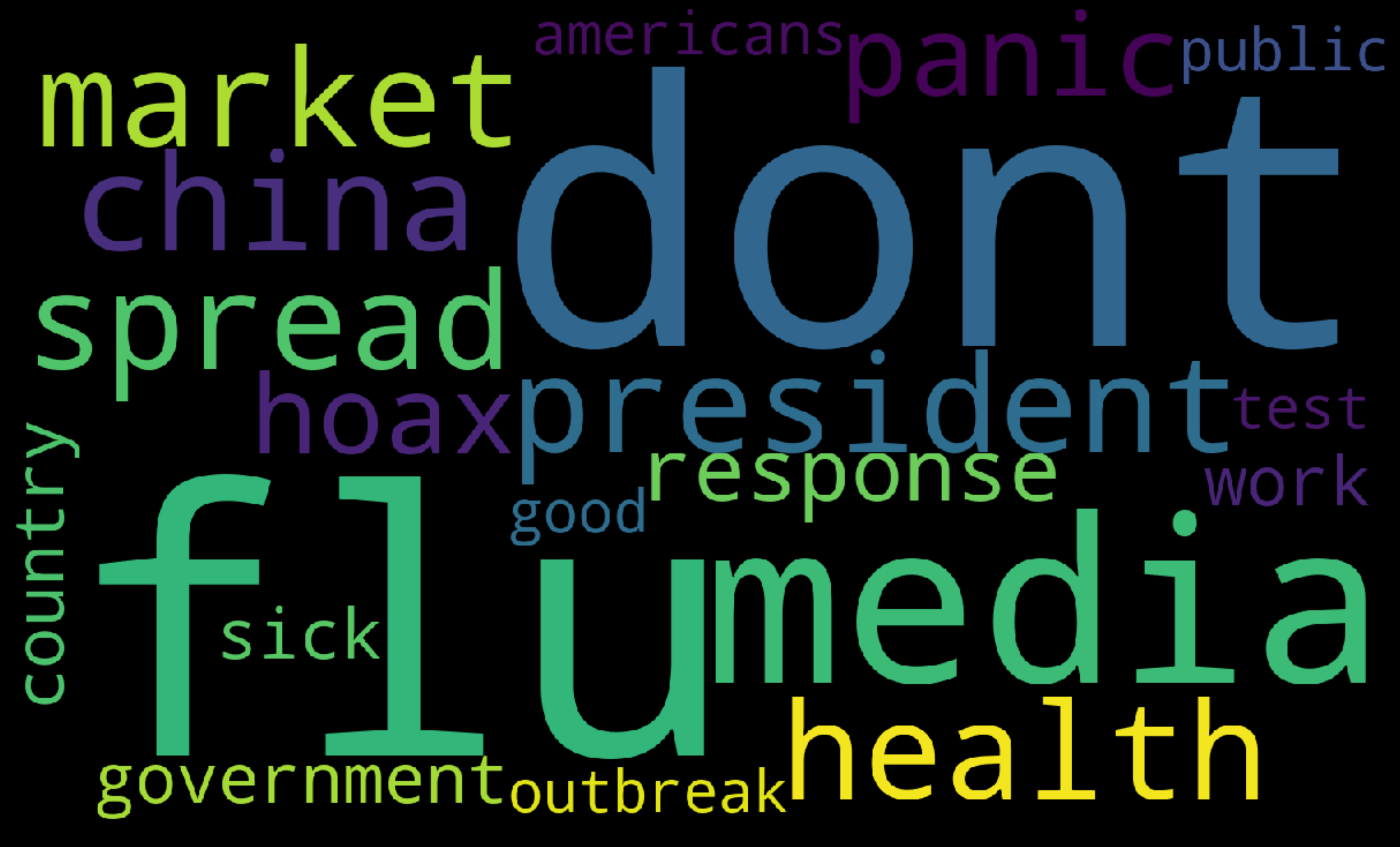}}
\subfigure[Denial]{
\label{enwcde}
\includegraphics[width=0.23\textwidth]{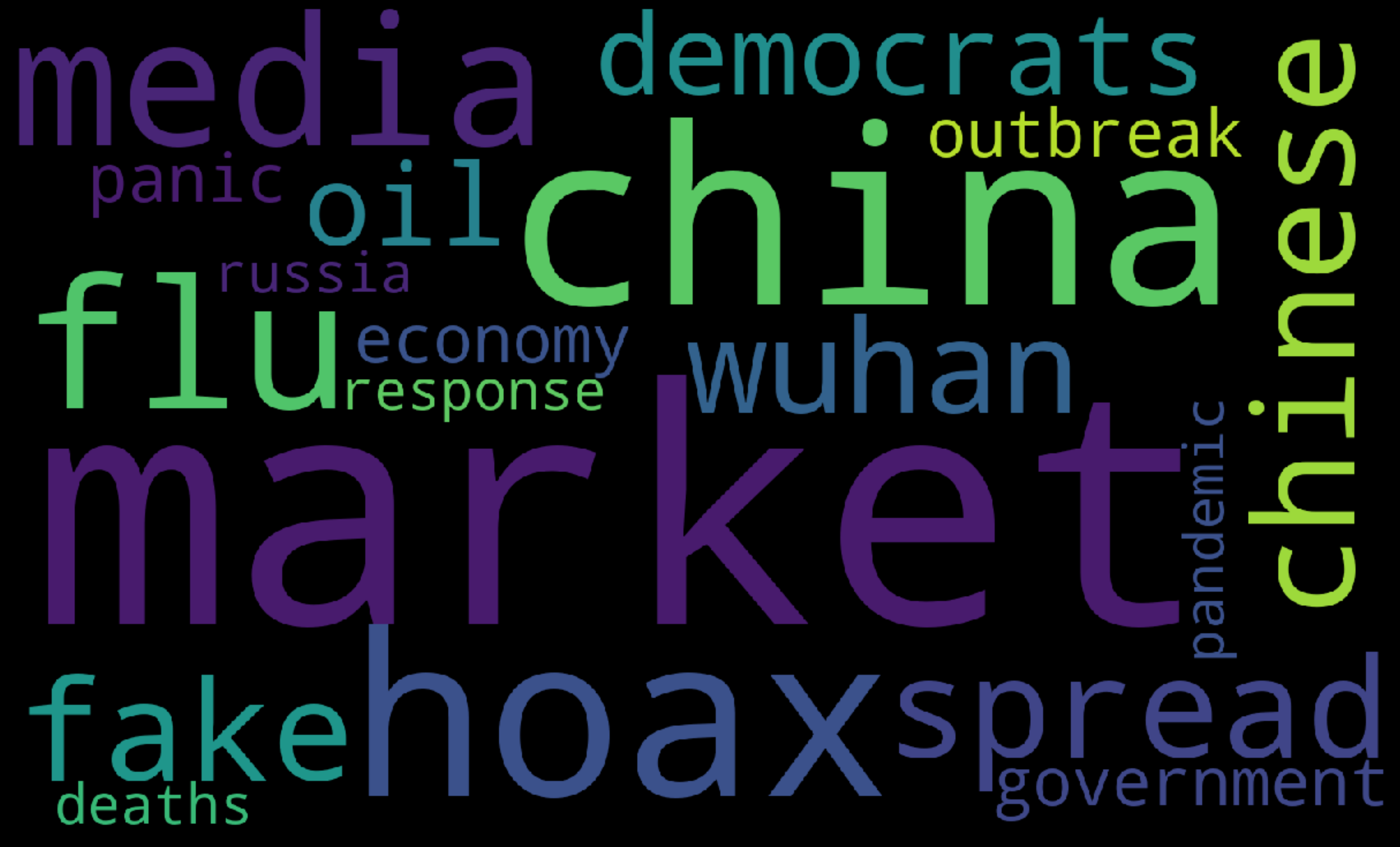}}
\subfigure[Official report]{
\label{enwcof}
\includegraphics[width=0.23\textwidth]{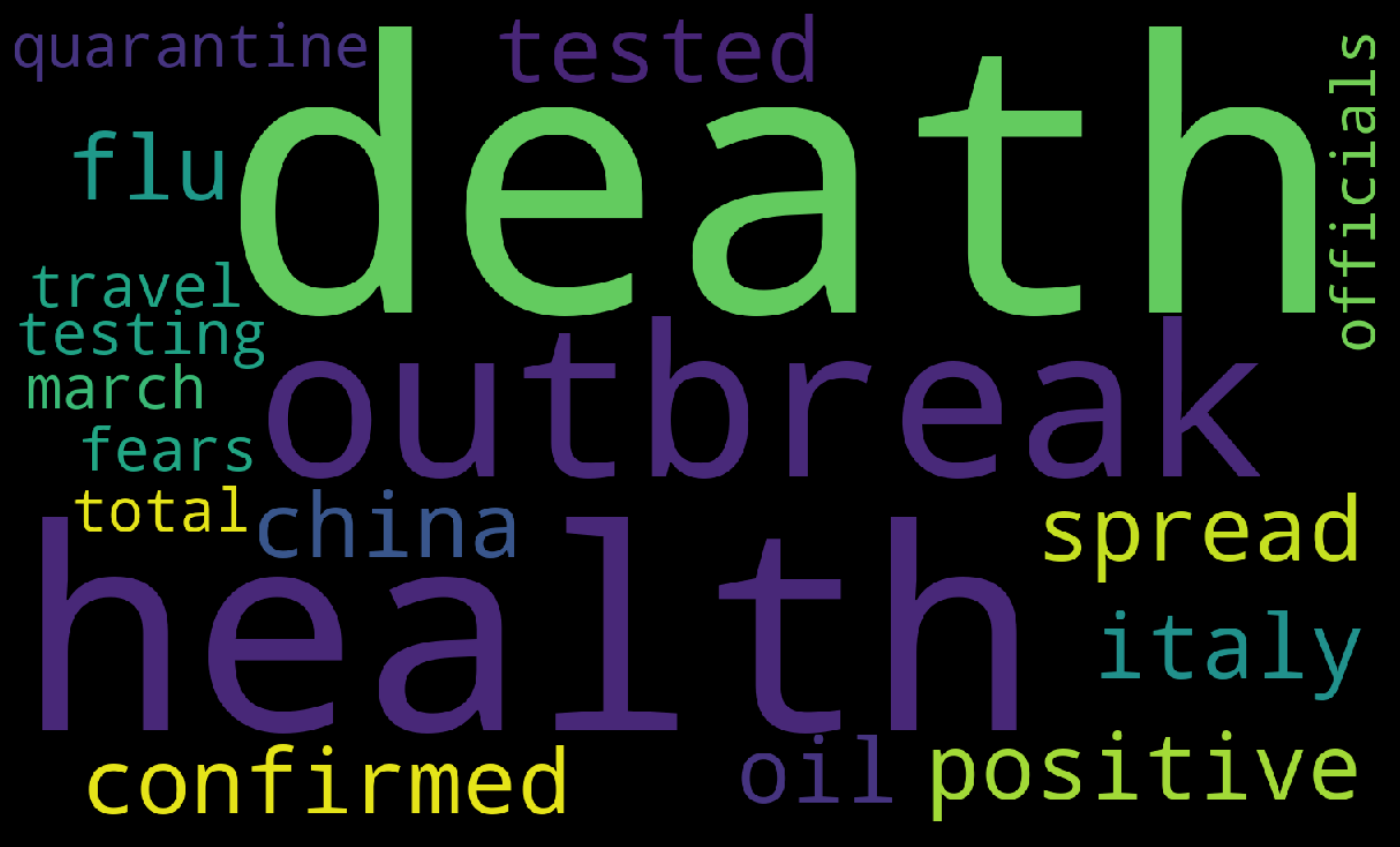}}
\subfigure[Joking]{
\label{enwcjo}
\includegraphics[width=0.23\textwidth]{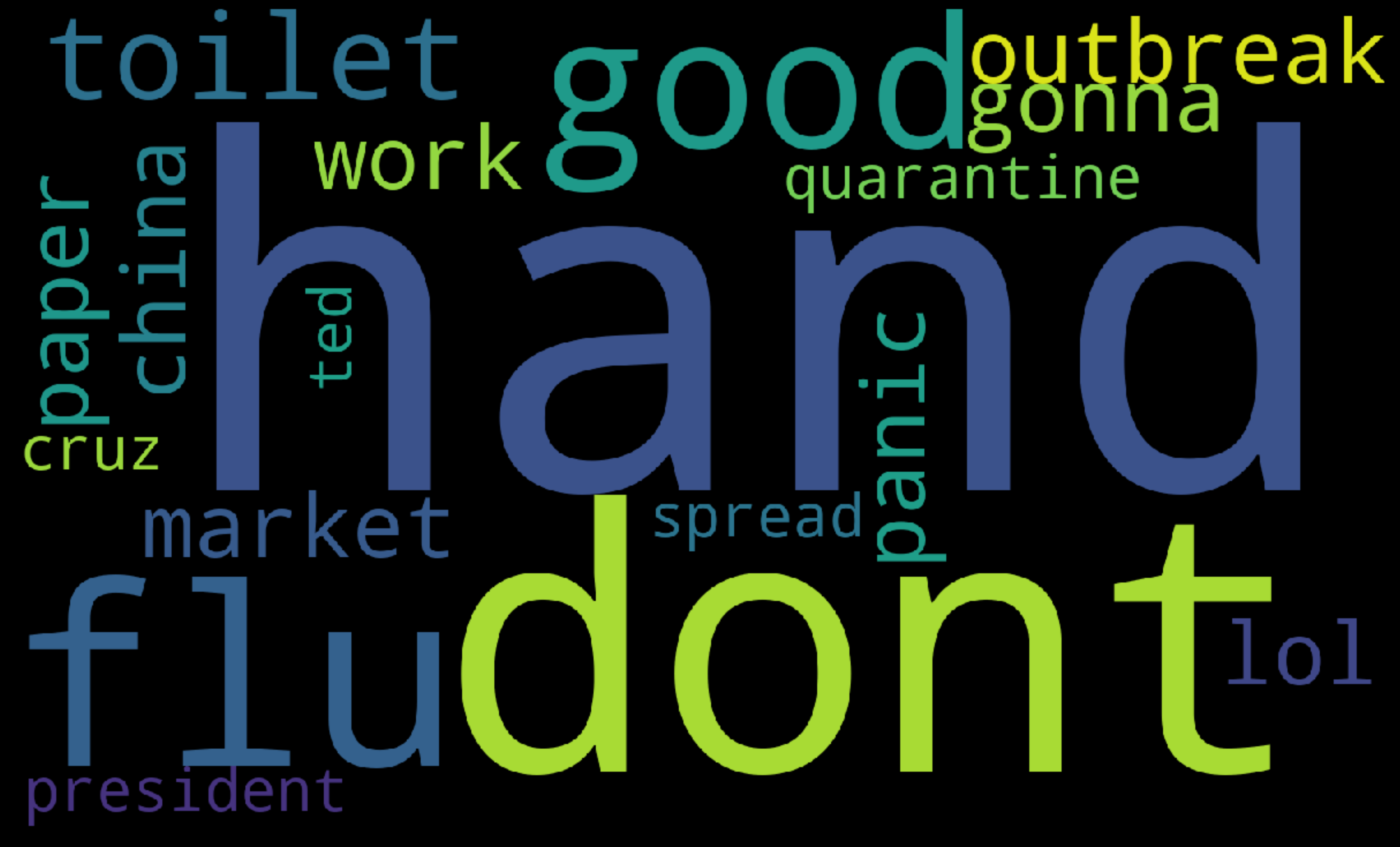}}
\caption{Hot words of each category for English tweets on March 9, 2020}
\label{fig:hwen}
\end{figure}

We present the hot words of the predicted English tweets for each category shown in Fig.~\ref{fig:hwen} on March 9, 2020. More representations of different languages will be provided in future. The class \emph{optimistic} is represented with hand washing and health, which means people should wash their hands frequently to keep health. The class \emph{thankful} is presented with Covid-19 testing, while the class \emph{empathetic} is shown with pray, hope, god and safe. The class \emph{pessimistic} is reflected with economy market, oil market and large number of death. These hot words are also suitable for the class \emph{anxious}. People felt \emph{sad} about a lot of death and confirmed cases and the lockdown of school. The class \emph{annoyed} is displayed with dont and flu while the class \emph{denial} is demonstrated with market and China since some people didn't believe the Covid-19 report of China. Overall, these hot words in each category can represent the sentiments to some extend.

\end{document}